 \documentclass{article}

\RequirePackage[latin1]{inputenc}     
 
\usepackage[english]{babel} 
\usepackage{amsmath} 
\usepackage{amsfonts} 
\usepackage{amssymb} 
\usepackage{amsthm}
\usepackage{xspace} 
\usepackage{latexsym} 
\usepackage{url} 
\usepackage{xspace} 
\usepackage{semantic}
\usepackage{cmll}
\usepackage{graphics,color}      
\usepackage{enumerate}        
\RequirePackage[latin1]{inputenc}

\newenvironment{restate-proposition}[2][{}]{\noindent\textbf{Proposition~{#2}}\;\textbf{#1}\  
}{\vskip 1em} 
 
\newenvironment{restate-theorem}[2][{}]{\noindent\textbf{Theorem~{#2}}\;\textbf{#1}\  
}{\vskip 1em} 
 
\newenvironment{restate-corollary}[2][{}]{\noindent\textbf{Corollary~{#2}}\;\textbf{#1}\  
}{\vskip 1em}

\newcommand{\hbra}{\noindent\hbox to \textwidth{\leaders\hrule height1.8mm depth-1.5mm\hfill}} 
\newcommand{\hket}{\noindent\hbox to \textwidth{\leaders\hrule height0.3mm\hfill}} 
\newcommand{\ratio}{.3}

\newtheorem{theorem}{Theorem} 
\newtheorem{lemma}{Lemma} 
\newtheorem{corollary}{Corollary} 
\newtheorem{proposition}{Proposition} 

\theoremstyle{definition}
\newtheorem{example}{Example}

\theoremstyle{remark}
\newtheorem{remark}{Remark} 

\newcommand{\Figbar}{{\center \rule{\hsize}{0.3mm}}}    %horizontal thiner line for figures 
 
\newcommand{\cl}[1]{{\cal #1}}          % \cl{R} to make R calligraphic 
\newcommand{\Gives}{\vdash}             % in a type judgement 
\newcommand{\IGives}{\vdash_{I}}        % intuitionistic provability
\newcommand{\AIGives}{\vdash_{{\it AI}}} %affine-intuitionistic provability
\newcommand{\CGives}{\vdash_{C}}        % affine-intuitionistic confluent provability

\newcommand{\arrow}{\rightarrow}        % right thin arrow 
\newcommand{\limp}{\multimap} %linear implication
\newcommand{\bang}{\oc} 
\newcommand{\limpe}[1]{\stackrel{#1}{\multimap}}
\newcommand{\hyp}[3]{#1:(#2, #3)}
\newcommand{\letm}[3]{{\sf let} \ ! #1 = #2 \ {\sf in} \ #3}    % modal let
\newcommand{\tertype}{{\bf 1}}
\newcommand{\behtype}{{\bf B}}
\newcommand{\Alt}{\mid} 
 
\newcommand{\csum}{\uplus}              %context sum
 
\newcommand{\infer}[2]{\begin{array}{c} #1 \\ \hline #2 \end{array}} 
\newcommand{\union}{\cup}               % union 
\newcommand{\dcl}{\downarrow}           % down closure 
\newcommand{\forget}[1]{\underline{#1}} % forgetful translation

\newcommand{\w}[1]{{\it #1}}    %To write in math style 

\newcommand{\qqs}[2]{\forall\, #1\;\: #2}

\newcommand{\s}[1]{{\sf #1}}    % sans-serif  

\newcommand{\act}[1]{\xrightarrow{#1}} %labelled actionlow %high

\newcommand{\set}[1]{\{#1\}}
\newcommand{\pst}[2]{{\sf pset}(#1,#2)}
\newcommand{\st}[2]{{\sf set}(#1,#2)}

\newcommand{\rgtype}[2]{{\it {\sf Reg}_{#1} #2}}

\newcommand{\get}[1]{{\sf get}(#1)}

\newcommand{\new}[2]{\nu #1 \ #2} 
 
\newcommand{\store}[2]{(#1 \leftarrow #2)}
\newcommand{\pstore}[2]{(#1 \Leftarrow #2)}
\newcommand{\regtype}[2]{{\sf Reg}_{#1} #2}

\newcommand{\upair}[2]{[#1,#2]}
\newcommand{\letb}[3]{\mathsf{let}\;\oc #1 = #2\;\mathsf{in}\;#3}

\newcommand{\vlt}[1]{{\cal V}(#1)}
\newcommand{\prs}[1]{{\cal P}(#1)}

\newenvironment{myexample}[1]{\begin{example}\hrulefill\\#1}{\hrule\end{example}}

\begin{document} 
 
\title{An Affine-Intuitionistic System of Types and Effects:
       Confluence and Termination}

\footnotetext[1]{Laboratoire PPS, Universit\'e Paris Diderot}
\footnotetext[2]{LIP - ENS-Lyon, Universit\'e de Lyon} 

\author{Roberto M. Amadio\footnotemark[1] \and  Patrick Baillot\footnotemark[2] \and Antoine Madet\footnotemark[1]}

\date{}

\maketitle 

\begin{abstract}
We present an affine-intuitionistic system of {\em types and effects} which
can be regarded as an extension  of Barber-Plotkin {\em Dual Intuitionistic Linear Logic}
to multi-threaded programs with effects. In the
system, dynamically generated values such as references or channels are
abstracted into a finite set of {\em regions}.  We introduce a discipline of
{\em region usage} that entails the {\em confluence} (and hence determinacy) of the
typable programs.  Further, we show that a discipline of region
{\em stratification} guarantees {\em termination}.
\end{abstract}

% \keywords{Linear logic, types and effects, confluence, termination}
% \terms{Languages, Theory}

\section{Introduction}\label{intro-sec}
There is a well-known connection between \emph{intuitionistic proofs} and {\em typed functional programs} that goes
under the name of {\em Curry-Howard} correspondence.  Following the
introduction of {\em linear logic} \cite{Girard87}, this correspondence has
been refined to include an explicit treatment of the process of data
duplication.  Various formalisations of these ideas have been proposed
in the literature (see, {\em e.g.}, \cite{BBPH93,Benton94,Plotkin93,MOTW95,Barber96}) 
and we will focus here in particular on Affine-Intuitionistic Logic and,
more precisely, on an {\em affine} version of Barber-Plotkin {\em Dual Intuitionistic
Linear Logic} (DILL) as described in \cite{Barber96}. 

In DILL, the operation of $\lambda$-abstraction is
always {\em affine}, {\em i.e.}, the formal parameter is used at most
once. The more general situation where the formal parameter has
multiple usages is handled through a constructor $`!'$ (read bang) 
marking values that can be duplicated and a destructor $\s{let}$ 
filtering them and effectively allowing their
duplication. Following this idea, {\em e.g.},  an intuitionistic judgement~(\ref{int-judg})
is translated into an affine-intuitionistic one~(\ref{aff-judg}) as follows:
\begin{align}
  \label{int-judg}
  y:A &\Gives \lambda x.x(xy): (A\arrow A) \arrow A\\
  \label{aff-judg}
  y:(\infty,A) &\Gives \lambda x.\letm{z}{x}{z\bang(z\bang y)}:\bang(\bang A\limp A\ )\limp A
\end{align}
We recall that in DILL the hypotheses are split in two zones according
to their {\em usage}.  Namely, one distinguishes between the {\em
  affine} hypotheses that can be used at most once and the {\em
  intuitionistic} ones that can be used arbitrarily many times. In our
formalisation, we will use $'1'$ for the former and $'\infty'$ for the
latter.

\subsection{Motivations}
Our purpose is to explore an {\em extension} of this connection to
{\em multi-threaded programs} with {\em effects}.  By extending the
connection, we mean in particular to design an affine-intuitionistic
type system that accounts for multi-threading and side effects 
and further to refine the system in order to guarantee 
confluence (and hence determinism)  and termination while 
preserving a reasonable expressive power.  By
multi-threaded program, we mean a program where distinct threads of
execution may be active at the same time (as it is typically the case
in concurrent programs) and by effect, we mean the possibility of
executing operations that modify the {\em state} of a system such as
reading/writing a reference or sending/receiving a message.
We stress that our aim is {\em not} to give a purely 
logical interpretation of multi-threading and side-effects 
but rather to apply {\em logical methods} to a multi-threaded 
programming language with side-effects.

\subsection{Contributions}
We will start by introducing a simple-minded extension of the purely
functional language with operators to run threads in parallel while
reading/writing the store which is loosely inspired by 
{\em concurrent} extensions of the ML programming language  
such as  \cite{GMP89} and \cite{Reppy91}
with an interaction mechanism
based on (asynchronous) {\em channel} communication.
In particular, we rely on 
an operator $\get{x}$ to read a value from an address (channel) $x$ and 
on two operators
$\st{x}{V}$ and $\pst{x}{V}$ to write a value $V$ into an address $x$,
in a volatile (value read is consumed) or persistent 
(value read is still available) way, respectively.

Following a rather standard practice 
(see, {\em e.g.}, \cite{LG88,TT97}), we suppose that
dynamically generated values such as channels or references  are
{\em abstracted} into a finite number of {\em regions}. This abstraction is
reflected in the type system where the type of an address {\em
depends} on the region with which the address is associated. Thus we 
write $\rgtype{r}{A}$ for the type of addresses containing values of
type $A$ and relating to the region $r$ of the store. Our first and probably
most difficult contribution, due to the interaction of the bang modality
$'\oc'$ with
regions, is to design a system where types and usages
are preserved by reduction. 

The resulting functional-concurrent typed language is
neither confluent nor terminating.  However, it turns out that there
are reasonable strategies to recover these properties.  The general
idea is that {\em confluence} can be recovered by introducing a proper
discipline of {\em region usage} while {\em termination} can be
recovered through a discipline of {\em region stratification}.

The notion of {\em region usage} is reminiscent 
of the one of {\em hypotheses usage} arising in affine-intuitionistic logic.
Specifically, we  distinguish the regions that can be used at most
once to write and at most once to read from those that can be used at
most once to write and arbitrarily many times to read. 
%The typing judgement then has the shape:
%\[\ldots r:(u,A)\ldots; \ldots x:(u',A')\ldots \Gives M:B \]

The notion of {\em region stratification} is based
on the idea that values stored in a region should only produce effects
on {\em smaller} regions.  The implementation of this idea requires a
substantial refinement of the type system that has to predict the {\em
effects} potentially generated by the evaluation of an expression.  This is 
where  {\em type and effect systems}, as introduced in
\cite{LG88}, come into play.

It turns out that the notions of region usage and region
stratification combine smoothly, leading to the definition of an
affine-intuitionistic system of types and effects. The system has
affine-intuitionistic logic as its functional core and it can be used
to guarantee the determinacy and termination of multi-threaded
programs with effects. We stress that 
the nature of our contribution is mainly methodological 
and that more theoretical and experimental work is needed to 
arrive at a usable  programming language. One promising direction
is to  add inductive data types and to extend the language to 
a synchronous/timed framework (cf. \cite{Amadio09,Boudol07}). 
In this framework, both confluence (determinism) and termination 
are valuable properties.

\subsection{Related Work}
Girard, through the introduction of {\em linear logic} \cite{Girard87}, 
has widely promoted a finer analysis of the {\em structural rules} of logic.
There have been various attempts at producing a functional programming
language based on these ideas and with a reasonably handy
syntax (see, {\em e.g.}, 
\cite{BBPH93,Benton94,Plotkin93,MOTW95,Barber96}).  The logical origin
of the notion of {\em usage} can be traced back to Girard's LU system
\cite{Girard91} and in particular it is adopted in the Barber-Plotkin system \cite{Barber96}
on which we build on.

A number of works on type systems for concurrent languages such as the
$\pi$-calculus have been inspired by linear logic even though in many
cases the exact relationships with logic are at best unclear even for the
fragment without side-effects. 
The conditions to guarantee {\em confluence} are inspired by
the work of  Kobayashi {\em et al.} \cite{KPT99} and one should expect
a comparable expressive power (see also  \cite{K02,IK05} for
much more elaborate notions of usage).

% \footnote{Let us also recall that Kobayashi
% {\em et al.} (see, {\em e.g.}, \cite{K02,IK05}) have produced type
% systems with a much more elaborate notion of {\em usage} than ours (a usage can
%be almost as complex as a CCS process) and shown that they can 
%guarantee a variety of properties of concurrent programs such as {\em absence of
%deadlock}.}

It is well known that intuitionistic logic is at the basis of typed
functional programming.  The {\em type and effect} system introduced
in \cite{LG88} is an enrichment of the intuitionistic system tracing
the effects of {\em imperative} higher-order programs acting on a {\em store}.  
The system has provided a successful static analysis tool
for the problem of {\em heap-memory deallocation} \cite{TT97}. 
More recently, this issue has been  revisited following the ideas of 
linear logic \cite{WW01,FMA06} .

The so called {\em reducibility candidates method} is probably the most
important technique to prove {\em termination} of typable higher-order
programs.  Extensions of the method to `functional fragments' of the
$\pi$-calculus have been proposed, {\em e.g.}, in \cite{YBH04,S06}.
Boudol \cite{Boudol07} has shown that a stratification of the regions
guarantees termination for a multi-threaded higher-order functional language 
with references and cooperative scheduling. 
Our formulation of the stratification discipline is actually based on \cite{Amadio09} 
which revisits and extends \cite{Boudol07}.

\subsection{Structure of the Paper}
Section \ref{aitype-sec} introduces an affine-intuitionistic system with regions
for a call-by-value functional-concurrent language.
Section \ref{confluence-sec} introduces a discipline of region usage that guarantees
confluence of the typable programs.
Section \ref{aitype-effect-sec} enriches the affine-intuitionistic system introduced
in Section \ref{aitype-sec} with a notion of effect which provides an upper bound
on the set of regions on which the evaluation of a term may produce effects.
Finally, Section \ref{termination-sec} describes a discipline of region stratification
that guarantees the termination of the typable programs.
Proofs of the main results are available in Appendix~\ref{appendix-proof}.

\section{An Affine-Intuitionistic Type System with Regions}\label{aitype-sec}
We introduce a typed functional-concurrent
programming language equipped with a call-by-value evaluation
strategy. 
The functional core of the language relies on Barber-Plotkin's
DILL.
In order to type the dynamically generated addresses of the
store, we introduce {\em regions} and suitable notions of {\em
  usages}. The related type system enjoys {\em weakening} and 
{\em substitution} and this leads to the expected properties of {\em type
preservation} and {\em progress}.

\subsection{Syntax: Programs}
Table \ref{syntax-terms} introduces the syntax of our programs.
\begin{table}
%{\footnotesize
\[
\begin{array}{r c l l}
\multicolumn{3}{l}{x,y,\ldots}                                            &\textrm{(Variables)} \\
V &::=& * \Alt x \Alt \lambda x.M \Alt \bang V                &\textrm{(Values)}\\
M &::=& V \Alt MM \Alt \bang M \Alt \letm{x}{M}{M} \Alt \new{x}{M} &\\
& &  \st{x}{V} \Alt \pst{x}{V} \Alt \get{x} \Alt (M\mid M) &\textrm{(Terms)}\\ 
S &::=& \store{x}{V} \Alt \pstore{x}{V} \Alt (S\mid S)   &\textrm{(Stores)} \\
P &::=& M \Alt S \Alt (P\mid P) \Alt \new{x}{P}          &\textrm{(Programs)}  \\
E &::=& [~]\Alt EM \Alt VE \Alt \bang E \Alt \letm{x}{E}{M}      &\textrm{(Evaluation Contexts)} \\
C &::=& [~] \Alt (C\mid P) \Alt (P\mid C) \Alt \new{x}{C}  &\textrm{(Static Contexts)}

\end{array}
\]%}
\caption{Syntax: programs}\label{syntax-terms}
\end{table}
We denote variables with $x,y,\ldots$, and  with $V$ the values
which are included in the category $M$ of terms. Stores are denoted by
$S$, and programs $P$ are combinations of terms and stores.  We comment on
the main operators of the language. $*$ is a constant inhabiting
the terminal type $\tertype$ (see below). $\lambda x.M$ is the {\em
affine} abstraction and $MM$ the application. $!$ marks values that
can be duplicated while $\letm{x}{M}{N}$ filters them and
allows their multiple usage in $N$. In $\new{x}{M}$ the operator $\nu$
generates a fresh address name $x$ whose scope is $M$. $\st{x}{V}$ and
$\pst{x}{V}$ write the value $V$ in a {\em volatile} address and a {\em
persistent} one, respectively, while $\get{x}$ fetches a value from
the address $x$ (either volatile or persistent). Finally $(M\mid N)$
evaluates in parallel $M$ and $N$.
Note that when writing either $\lambda x.M$, or $\new{x}{M}$, or 
$\letm{x}{N}{M}$ the
variable $x$ is bound in $M$. 
As usual, we abbreviate $(\lambda z.N)M$ with $M;N$, where $z$ is not free in $N$.
{\em Evaluation contexts} $E$ follow a {\em call-by-value} discipline.
{\em Static contexts} $C$ are composed of parallel composition and $\nu$'s.
Note that stores can only appear in a static context. Thus 
$M=V(\st{x}{V'};V'')$ is a legal term while  $M'=V(V''\mid \store{x}{V})$ is not.

\subsection{Operational Semantics}
Table \ref{op-sem} describes the operational semantics of our language.
\begin{table}[h]
%{\footnotesize
\[
\begin{array}{rcll}

P\mid P' &\equiv &P'\mid P                                &\textrm{(Commut.)} \\
(P\mid P')\mid P'' &\equiv &P \mid (P' \mid P'')          &\textrm{(Assoc.)} \\
\new{x}{P}\mid P' &\equiv &\new{x}{(P\mid P')} \quad x\notin \w{FV}(P')  &(\nu_{\mid}) \\
E[\new{x}{M}] &\equiv &\new{x}{E[M]} \quad x\notin \w{FV}(E)            &(\nu_{E}) 
\end{array}
\]
\[
\begin{array}{rcl}
E[(\lambda x.M)V]  &\arrow &E[[V/x]M] \\
E[\letm{x}{\bang V}{M}] &\arrow &E[[V/x]M] \\
E[\st{x}{V}]       &\arrow &E[*]\mid \store{x}{V} \\
E[\pst{x}{V}]      &\arrow &E[*]\mid \pstore{x}{V} \\
E[\get{x}] \mid \store{x}{V} &\arrow &E[V] \\
E[\get{x}] \mid \pstore{x}{\bang V} &\arrow &E[\bang V] \mid \pstore{x}{\bang V}

\end{array}
\]%}
\caption{Operational semantics} \label{op-sem}
\end{table}
Programs are considered up to a {\em structural equivalence} $\equiv$ which
is the least equivalence relation preserved by  static contexts,
and which contains the equations for  $\alpha$-renaming, 
for the commutativity and associativity  of parallel composition,
for enlarging the scope of the $\nu$ operators to parallel programs,
and for extracting the $\nu$ from an evaluation context.
We use the notation $[V/x]$ for the substitution of the value $V$
for the variable $x$.
The reduction rules apply modulo structural equivalence 
and in a static context $C$.
\begin{myexample}
  The programs (\ref{program1}) and (\ref{program2}) are structurally equivalent (up
to some renaming):
\begin{gather}
  \label{program1}
  ((\new{x}{\lambda y.M})(\new{x'}{\lambda x'.M'}))V \mid P\\
  \label{program2}
  \new{x}{\new{x'}{((\lambda y.M)(\lambda y'.M'))V}} \mid P  
\end{gather}
This transformation
exposes the term  $E[(\lambda y.M)(\lambda y'.M')]$ in the 
static context $C=\new{x}{\new{x'}{[~]}} \mid P$, where the
evaluation context $E$ is $[~]V$.
\vspace{4pt}
\end{myexample}
In the sequel we consider the transitive closure of the relation
defined by Table 2, also  denoted $\rightarrow$.

\begin{remark}
Notice that the \textsf{let} rule and the \textsf{get} rule on a
persistent store act similarly in the sense that they require the
value being duplicated to be marked with a bang, while the affine
$\beta$ rule and the \textsf{get} rule on a volatile store allow to
manipulate affine values.  
\end{remark}

\subsection{Syntax: Types and Contexts}
Table \ref{syntax-types} introduces the syntax of types and contexts.
\begin{table}[h]
%{\footnotesize
\[
\begin{array}{r c l l}
\multicolumn{3}{l}{r,r',\ldots}                                           &\textrm{(Regions)} \\
\alpha &::=& \behtype \Alt A                             &\textrm{(Types)} \\
A &::=& \tertype \Alt A\limp \alpha \Alt \bang A \Alt \rgtype{r}{A}   &\textrm{(Value-types)} \\
\Gamma &::=& \hyp{x_{1}}{u_{1}}{A_{1}},\ldots,\hyp{x_{n}}{u_{n}}{A_{n}}  &\textrm{(Contexts)} \\
R &::=& \hyp{r_{1}}{U_{1}}{A_{1}},\ldots,\hyp{r_{n}}{U_{n}}{A_{n}}  &\textrm{(Region contexts)} \\

\end{array}
\]%}
\caption{Syntax: types and contexts}\label{syntax-types}
\end{table}
We denote regions with $r,r',\ldots$ and we suppose a region $r$
is either {\em volatile} ($\vlt{r}$) or {\em persistent} ($\prs{r}$).
Types are denoted with $\alpha,\alpha',\ldots$. Note that we distinguish a
special behaviour type $\behtype$ which is given to the entities of
the language which are
not supposed to return a value  (such as a store or several values in parallel)
while types of entities that may return a value are denoted with $A$.
Among the types $A$, we distinguish a terminal type $\tertype$, 
an affine functional type $A \limp B$, the type $\bang A$ of terms of type $A$
that can be duplicated, and the type $\rgtype{r}{A}$ of 
addresses containing values of type $A$
and related to the region $r$. Hereby types may depend on regions.

Before commenting on variable and region contexts, we need to define the notion of {\em usage}.
To this end, it is convenient to introduce a set with three values 
$\set{0,1,\infty}$ and a {\em partial} binary operation $\csum$ such
that 
$$
\begin{array}{r c l}
%\hline
x\csum 0 &=& x\\
0\csum x &=& x\\
\infty \csum \infty &=& \infty\\
%\hline
\end{array}
$$
and which is undefined otherwise.

We denote with $u$ a {\em variable usage} and assume that $u$ is either $1$ (a variable to be used
at most once) or $\infty$ (a variable that can be used arbitrarily many times).
Then a variable context (or simply a context) $\Gamma$ has the  shape:
$$\hyp{x_{1}}{u_{1}}{A_{1}},\ldots,\hyp{x_{n}}{u_{n}}{A_{n}}$$
where $x_i$ are distinct variables, $u_i\in \set{1,\infty}$ and $A_i$ are types of
terms that may return a result. Writing $\hyp{x}{u}{A}$ means that the variable 
$x$ ranges on values of type $A$ and can be used according to $u$.
We  write $\w{dom}(\Gamma)$ for the set $\set{x_1,\ldots,x_n}$ of
variables where the context is defined.
The sum on usages is extended to contexts component-wise.
In particular, if  $x:(u_1,A)\in \Gamma_1$ and $x:(u_2,A)\in \Gamma_2$ then
$x:(u_1\csum u_2,A)\in (\Gamma_1\csum \Gamma_2)$ only if 
$u_1\csum u_2$ is defined.
\begin{myexample}
  One may check that the sum:
  $$(x:(1,A),y:(\infty,B)) \uplus (y:(\infty,B),z:(1,C))$$
  is equal to
  $$x:(1,A),y:(\infty,B),z:(1,C)$$
  whereas these two are not defined:
  \begin{align*}
    (x:(1,A),y:(\infty,B)) &\uplus y:(1,B)\\
    (x:(1,A),y:(1,B) &\uplus y:(1,B)
  \end{align*}
\end{myexample}

We are going to associate a usage with regions too, but in this case a usage
will be a two dimensional vector because we want to be able to distinguish
write and read usages. We denote with $U$ an element of one of the following
three sets of usages:
$$
\begin{array}{ c c c }
%\hline&&\\
\set{\upair{\infty}{\infty}} &
\set{\upair{1}{\infty},\upair{0}{\infty}}
&\set{\upair{0}{0},\upair{1}{0},\upair{0}{1},\upair{1}{1}}\\
%& &\\
%\hline
\end{array}
$$
where by convention we reserve the first component to describe the
write usage and the second for the read usage. Thus a region with usage
$\upair{1}{\infty}$ should be written at most once while it can be 
read arbitrarily many times. 

The addition $U_1\csum U_2$ is defined provided that:
\begin{enumerate}[(a)]
\item $U_1$ and $U_2$ are in the same set of usages
\item the component-wise addition is defined
\end{enumerate}
\begin{myexample} If $U_1=\upair{\infty}{\infty}$ and $U_2=\upair{0}{\infty}$ then the sum is undefined
because $U_1$ and $U_2$ are not in the same set while if $U_1=\upair{1}{\infty}$ and
$U_2=\upair{1}{\infty}$ then the sum is undefined because $1\csum 1$
is undefined.
\vspace{4pt}
\end{myexample}
Note that in each set of usages there is
a {\em neutral} usage $U_0$ such that $U_0\csum U=U$ for all $U$ in the same set.

A region context $R$ has the shape:
$$
\hyp{r_{1}}{U_{1}}{A_{1}},\ldots,\hyp{r_{n}}{U_{n}}{A_{n}}
$$
where $r_i$ are distinct regions, $U_i$ are usages in the sense just defined, 
and $A_i$ are value-types.
The typing system will additionally guarantee that whenever we use a type 
$\rgtype{r}{A}$ the region context  contains 
an hypothesis $\hyp{r}{U}{A}$ for some $U$.
Intuitively, writing $\hyp{r}{U}{A}$ means that addresses related to region
$r$ contain values of type $A$ and that they 
can be used according to the usage $U$.
We  write $\w{dom}(R)$ for the set $\set{r_1,\ldots,r_n}$ of the regions
where the region context is defined.
As for contexts, the sum on usages is extended to region contexts
component-wise.
In particular,  if \mbox{$r:(U_1,A)\in R_1$} and $r:(U_2,A)\in R_2$ then
\mbox{$r:(U_1\csum U_2,A)\in (R_1\csum R_2)$} only if 
$U_1\csum U_2$ is defined. 
Moreover, for $(R_1\csum R_2)$ to be defined we require that $\w{dom}(R_1)=\w{dom}(R_2)$.
There is no loss of generality in this hypothesis because if, say, 
$\hyp{r}{U}{A}\in R_1$ and $r\notin\w{dom}(R_2)$ then we can always add 
$\hyp{r}{U_{0}}{A}$ to $R_2$ where $U_0$ is the neutral usage of the set to which
$U$ belongs (this is left implicit in the typing rules).

\begin{myexample}
  One may check that the sum:
  \begin{align*}
    (&\hyp{r_1}{\upair{1}{\infty}}{A},\hyp{r_2}{\upair{0}{1}}{B})\\
    &\qquad\qquad\qquad \uplus (\hyp{r_1}{\upair{0}{\infty}}{A},\hyp{r_2}{\upair{1}{0}}{B})
  \end{align*}
  is equal to
  $$\hyp{r_1}{\upair{1}{\infty}}{A},\hyp{r_2}{\upair{1}{1}}{B}$$
  whereas these two are not defined:
  \begin{align*}
    (R,\hyp{r}{\upair{1}{\infty}}{B})
    &\uplus (R,\hyp{r}{\upair{1}{\infty}}{B})\\
    (R,\hyp{r}{\upair{0}{\infty}}{B})
    &\uplus (R,\hyp{r}{\upair{1}{0}}{B})
  \end{align*}
\end{myexample}

\subsection{Affine-Intuitionistic Type System with Regions}
Because types depend on regions, we have to be careful in stating in 
Table \ref{type-cxt-unstratified}
when a region-context and a type are compatible ($R\dcl \alpha$),
when a region context is well-formed ($R\Gives$), 
when a type is well-formed in a region context (\mbox{$R\Gives \alpha$}) 
and when a context is well-formed in a region context (\mbox{$R\Gives \Gamma$}).

A more informal way to express the condition is to say that a
judgement $\hyp{r_{1}}{U_{1}}{A_{1}},\ldots,\hyp{r_{n}}{U_{n}}{A_{n}}
\Gives \alpha$ is well formed provided that: 
\begin{enumerate}[(a)]
\item all the region names
occurring in the types $A_1,\ldots,A_n,\alpha$ belong to the set
$\set{r_1,\ldots,r_n}$ 
\item all types of the shape
$\rgtype{r_{i}}{B}$ with $i\in \set{1,\ldots,n}$ and occurring in the
types $A_1,\ldots,A_n,\alpha$ are such that \mbox{$B=A_i$}.
\end{enumerate}
\begin{myexample}
One may verify that $$\hyp{r}{U}{\tertype \limp \tertype}
\Gives \rgtype{r}{(\tertype \limp \tertype)}$$ can be derived while
these judgements cannot:
\begin{align*}
\hyp{r}{U}{\tertype} &\Gives \rgtype{r}{(\tertype \limp \tertype)}\\
\hyp{r}{U}{\rgtype{r}{\tertype}} &\Gives \tertype
\end{align*}
\end{myexample}

\begin{table}[h]
%{\footnotesize
\[
\begin{array}{c c}

\inference
{}
{R\dcl \tertype}
&

\inference
{}
{R\dcl \behtype}
\\\\

\inference
{R\dcl A & R\dcl \alpha}
{R\dcl (A\limp \alpha)}

&

\inference
{\hyp{r}{U}{A}\in R}
{R\dcl \rgtype{r}{A}}

\\ \\

\inference
{\qqs{\hyp{r}{U}{A} \in R} & R\dcl A}
{R\Gives}

&

\inference
{R\Gives & R\dcl \alpha}
{R\Gives \alpha} 

\\ \\

\multicolumn{2}{c}{\inference
{\qqs{\hyp{x}{u}{A}\in \Gamma} & R\Gives A }
{R\Gives \Gamma}}

\end{array}
\]%}
\caption{Type and context formation rules (unstratified)}\label{type-cxt-unstratified}
\end{table}
Next, Table \ref{air-system} introduces an affine-intuitionistic type system 
{\em with regions} whose basic judgement
$$R;\Gamma \Gives P:\alpha$$ attributes a type $\alpha$ to the program $P$
in the region context $R$ and the context $\Gamma$.
Here and in the following we omit the rule for typing a program
$(S\mid P)$ which is symmetric to the one for the program $(P\mid S)$.

The formulation of the so called {\em promotion rule}, {\em i.e.},
the rule that introduces the $`!'$ operator, requires some
care. In particular, we notice that its formulation relies on 
the predicates \w{aff} (affine) and  \w{saff} (strongly affine) 
on contexts and region contexts which we define below.
The intuition is that terms whose typing depends on affine
(region) contexts should {\em not} be duplicated, 
{\em i.e.}, should not be `marked' with \mbox{a $\bang$}.
Formally, we write $\w{aff}(\hyp{x}{u}{A})$ if $u=1$. 
We also write $\w{aff}(\hyp{r}{\upair{v}{v'}}{A})$ 
if either $1\in \set{v,v'}$ or ($\vlt{r}$ and $v'\neq 0$). 
Moreover, we write $\w{aff}(R;\Gamma)$ (respectively $\w{saff}(R;\Gamma)$)
if the predicate $\w{aff}$ holds for at least one of (respectively for all) the hypotheses in
$R;\Gamma$. 

\begin{remark}
Notice that  we regard the hypothesis $\hyp{r}{\upair{v}{v'}}{A}$ 
as affine  if either it contains the information that we can read or write 
in $r$ at most once or if $r$ is a volatile region from which we can read. 
The reason for the second condition is that a volatile region
may contain data that should be used at most once. 
For instance, assuming 
$\vlt{r}$, 
$R=\hyp{r}{\upair{\infty}{\infty}}{A}$, and 
$\Gamma=\hyp{x}{\infty}{\rgtype{r}{A}}$, we can derive
\mbox{$R;\Gamma \Gives \get{x}:A$}. However, we
should {\em not} derive \mbox{$R;\Gamma \Gives \bang \get{x}: \bang A$} for
otherwise the crucial subject reduction property (Theorem~\ref{sub-red-thm})
may be compromised.
\end{remark}

Finally, we remark that in the conclusion of the {\em promotion rule} 
we may weaken the (region) context with a strongly affine (region) context.
This is essential to obtain the following  weakening property.
\begin{lemma}[weakening]
If $R;\Gamma \Gives P:\alpha$ and $R\csum R'\Gives \Gamma\csum \Gamma'$ then
$R\csum R';\Gamma\csum \Gamma'\Gives P:\alpha$.
\end{lemma}

\begin{table}
%{\footnotesize
\[
\begin{array}{c}

\inference
{R\Gives \Gamma & \hyp{x}{u}{A}\in \Gamma}
{R;\Gamma \Gives x:A}

\qquad

\inference
{R\Gives \Gamma}
{R;\Gamma \Gives *:\tertype} \\ \\

\inference
{R;\Gamma,\hyp{x}{1}{A} \Gives M:\alpha}
{R;\Gamma \Gives \lambda x.M:(A \limp \alpha)}

\qquad

\inference
{R_1;\Gamma_1\Gives M:(A\limp \alpha)\\
R_2;\Gamma_2\Gives N:A}
{R_1\csum R_2;\Gamma_1\csum \Gamma_2 \Gives MN:\alpha} \\ \\ 

\inference
{R\csum R'\Gives (\Gamma\csum \Gamma') & \w{saff}(R';\Gamma')\\
R;\Gamma \Gives M:A & \neg \w{aff}(R;\Gamma) 
%\hyp{x}{u}{B}\in \Gamma' \textrm{ implies }u=1
}
{R\csum R';\Gamma \csum \Gamma' \Gives \bang M:\bang A}

\qquad

\inference{
R_1;\Gamma_1\Gives M:\bang A\\ 
R_2;\Gamma_2,\hyp{x}{\infty}{A} \Gives N:\alpha
}
{R_1\csum R_2;\Gamma_1\csum \Gamma_2 \Gives \letm{x}{M}{N}:\alpha} \\ \\

\inference{R;\Gamma,\hyp{x}{u}{\rgtype{r}{A}} \Gives P:\alpha}
{R;\Gamma \Gives \new{x}{P}:\alpha}

\qquad
\inference{
R\Gives \Gamma & \hyp{x}{u}{\rgtype{r}{A}}\in \Gamma  \\
\hyp{r}{\upair{v}{v'}}{A} \in R & v' \neq 0 
}
{R;\Gamma \Gives \get{x}:A} \\ \\

\inference{
\Gamma= \hyp{x}{u}{\rgtype{r}{A}} \csum \Gamma' & \vlt{r}\\
R = \hyp{r}{\upair{v}{v'}}{A} \csum R' & v \neq 0 \\
R\Gives \Gamma & R';\Gamma'\Gives V:A
}
{R;\Gamma \Gives \st{x}{V}:\tertype} 

\qquad

\inference{
\Gamma= \hyp{x}{u}{\rgtype{r}{\bang A}} \csum \Gamma'& \prs{r}\\
R = \hyp{r}{\upair{v}{v'}}{\bang A} \csum R' & v \neq 0 \\
R\Gives \Gamma & R';\Gamma'\Gives V:\bang A
}
{R;\Gamma \Gives \pst{x}{V}:\tertype} \\ \\

\inference{
\Gamma= \hyp{x}{u}{\rgtype{r}{A}} \csum \Gamma'& \vlt{r}\\
R = \hyp{r}{\upair{v}{v'}}{A} \csum R' & v \neq 0 \\
R\Gives \Gamma & R';\Gamma'\Gives V:A
}
{R;\Gamma \Gives \store{x}{V}:\behtype} 

\qquad

\inference{
\Gamma= \hyp{x}{u}{\rgtype{r}{\bang A}} \csum \Gamma' & \prs{r}\\
R = \hyp{r}{\upair{v}{v'}}{\bang A} \csum R' & v \neq 0 \\
R\Gives \Gamma & R';\Gamma'\Gives V:\bang A
}
{R;\Gamma \Gives \pstore{x}{V}:\behtype} \\ \\

\inference{R_1;\Gamma_1\Gives P:\alpha& R_2;\Gamma_2\Gives S:\behtype}
{R_1\csum R_2; \Gamma_1\csum \Gamma_2 \Gives (P \mid S):\alpha}

\qquad

\inference{R_i;\Gamma_i\Gives P_i:\alpha_i& P_i \textrm{ not a store } i=1,2}
{R_1\csum R_2; \Gamma_1\csum \Gamma_2 \Gives (P_1 \mid P_2):\behtype}

\end{array}
\]%}
\caption{An affine-intuitionistic type system with regions}\label{air-system}
\end{table}

Then we see how our type system applies to some program examples.
\begin{myexample}
Let $R=\hyp{r}{\upair{1}{1}}{\tertype}$ and
$$M=\lambda x.\letm{x}{x}{\get{x} \mid \st{x}{*}}$$
We check that: $$R;\_ \Gives M: \bang \regtype{r}{\tertype} \limp \behtype$$
By the rule for affine implication, this reduces to:
$$R;\hyp{x}{1}{\bang \regtype{r}{\tertype}} \Gives \letm{x}{x}{\get{x} \mid \st{x}{*}}: \behtype$$
If we define $R_0=\hyp{r}{\upair{0}{0}}{\tertype}$, then by the rule for the
{\sf let} we reduce to:
$$R_0; \hyp{x}{1}{\bang \regtype{r}{\tertype}} \Gives x: \bang \regtype{r}{\tertype}$$
and 
$$R; \hyp{x}{\infty}{\regtype{r}{\tertype}} \Gives \get{x} \mid \st{x}{*} : \behtype$$
The former is an axiom while the latter is derived from:
$$\hyp{r}{\upair{0}{1}}{\tertype};  
\hyp{x}{\infty}{\regtype{r}{\tertype}}
\Gives \get{x}:\tertype$$
and
$$\hyp{r}{\upair{1}{0}}{\tertype};  
\hyp{x}{\infty}{\regtype{r}{\tertype}} 
\Gives \st{x}{*}:\tertype$$
Note that we can actually apply the function $M$ to a value $\bang y$ which 
is typed using the promotion rule as follows:
\[
\inference{R_0;\hyp{y}{\infty}{\regtype{r}{\tertype}} \Gives y: \regtype{r}{\tertype}}
{R_0;\hyp{y}{\infty}{\regtype{r}{\tertype}} \Gives \bang y: \bang \regtype{r}{\tertype}}
\]
We remark that the region context and the context play two
different roles: the context counts the number of occurrences of a variable
while the region context counts the number of read-write effects. 
In our example, the variable $x$ occurs several times but we can be sure
that there will be at most one read and at most one write in the related
region.\vspace{4pt}
\end{myexample}

\begin{myexample}
We consider a {\em functional} $$M=\lambda f.\lambda f'.\new{y}{(fy\mid  f'y)}$$
which can be given the type 
$$(\rgtype{r}{\tertype}\limp\tertype) \limp 
 (\rgtype{r}{\tertype}\limp\tertype) \limp \behtype$$ in a region
context $R=\hyp{r}{\upair{0}{0}}{\tertype}$. We can apply $M$ to 
the functions 
$$
\begin{array}{c @{\; \textrm{and} \;} c}
V_1=\lambda x.\get{x} & V_2=\lambda x.\st{x}{*}  
\end{array}
$$
which
have the appropriate types in the compatible region contexts 
$R'= \hyp{r}{\upair{0}{1}}{\tertype}$ and 
$R''= \hyp{r}{\upair{1}{0}}{\tertype}$, respectively.
Such {\em affine} usages would not be compatible with an
intuitionistic implication as in this case
one has to {\em promote} (put a $\bang$ in front of) 
$V_1$ and $V_2$ before passing them as arguments.\vspace{4pt}
\end{myexample}

As in Barber-Plotkin system \cite{Barber96}, the 
substitution lemma comes in two flavours:
\begin{lemma}[substitution]\label{sub-lemma}
 Affine substitution~(\ref{aff-sub}) and intuitionistic
 substitution~(\ref{int-sub}) preserve typing:
  \begin{enumerate}[(a)]
  \item\label{aff-sub} If $R;\Gamma,\hyp{x}{1}{A} \Gives M:\alpha$,
$R';\Gamma'\Gives V: A$, and 
$R\csum R' \Gives \Gamma\csum \Gamma'$ then
$R\csum R';\Gamma\csum \Gamma'\Gives [V/x]M:\alpha$.
\item\label{int-sub} If $R;\Gamma,\hyp{x}{\infty}{A} \Gives M:\alpha$,
$R';\Gamma'\Gives \bang V:\bang A$, and 
$R\csum R' \Gives \Gamma\csum \Gamma'$ then
$R\csum R';\Gamma\csum \Gamma'\Gives [V/x]M:\alpha$.
  \end{enumerate}
\end{lemma}

We rely on Lemma \ref{sub-lemma} to show that the basic
reduction rules in Table \ref{op-sem} preserve typing.
Then, observing that typing is invariant under structural
equivalence, we can lift the property  to the reduction relation 
which is generated by the basic reduction rules.

\begin{theorem}[subject reduction]\label{sub-red-thm}
If $R;\Gamma \Gives P:\alpha$ and $P\arrow P'$ then
$R;\Gamma \Gives P':\alpha$. 
\end{theorem}

In our formalism, a {\em closed} program is a program whose only free variables
have region types (as in, say, the $\pi$-calculus). 
For {\em closed} programs one can state a {\em progress property}
saying that if a program cannot progress then, up to structural equivalence,
every thread is either a value or a term of the shape $E[\get{x}]$ and there
is no store in parallel of the shape $\store{x}{V}$ or $\pstore{x}{V}$. In particular,
we notice that a {\em closed} value of type $!A$ must have the shape $!V$ so
that in well-typed closed programs such as $\letm{x}{V}{M}$ or
\mbox{$E[\get{x}]\mid \pstore{x}{V}$}, $V$ is guaranteed to have the shape $!V$ 
required by the operational semantics in Table \ref{op-sem}.
\begin{proposition}[progress]
Suppose $P$ is a closed typable program which cannot reduce. Then
$P$ is structurally equivalent to a program
\[
\nu x_1,\ldots,x_m \ (M_1 \mid \cdots \mid M_n \mid S_1 \mid \cdots
\mid S_p)\quad m,n,p\geq 0
\]
where $M_i$ is either a value or can be uniquely decomposed as 
a term $E[\get{y}]$ such that no value is associated with the address
$y$ in the stores $S_1,\ldots,S_p$.
\end{proposition}

\section{Confluence}\label{confluence-sec}
In our language, each thread evaluates deterministically
according to a call-by-value evaluation strategy.
The only source of non-determinism comes from a concurrent
access to the memory. More specifically, we may have a non-deterministic
program if several values are stored at the same address as in the following
examples (note that we cannot type a program where values are stored at an address both in a persistent and a volatile way):
\begin{align}\label{value-race1}
\get{x} \mid \pstore{x}{V_1} \mid \pstore{x}{V_2} \\
\label{value-race2}
\get{x} \mid \store{x}{V_1} \mid \store{x}{V_2}
\end{align}
or if there is a race condition on a volatile address as in the following
example:
\begin{equation}\label{volatile-store-race}
E_1[\get{x}] \mid E_2[\get{x}] \mid \store{x}{V}
\end{equation}
On the other hand, a race condition on a persistent address such as:
\begin{equation}\label{persistent-store-race}
E_1[\get{x}] \mid E_2[\get{x}] \mid \pstore{x}{V}
\end{equation}
does not compromise determinism because the two possible reductions commute.
We can rule out the problematic situations~\eqref{value-race1},~\eqref{value-race2} 
and (\ref{volatile-store-race}),
if:
\begin{enumerate}[(a)]
\item we remove from our system the region usage
$\upair{\infty}{\infty}$
\item we restrict
the usages of volatile stores to those 
in which there is at most one read effect (hence the set
$\set{\upair{1}{1},\upair{1}{0},\upair{0}{1},\upair{0}{0}}$)
\end{enumerate}
To this end, we add a condition $v'\neq \infty$ to the typing rules
for volatile stores $\st{x}{V}$ and $\store{x}{V}$ as specified in
Table \ref{rules-confluence}.  
\begin{table}[h]
%{\footnotesize
\[
\begin{array}{c}

U\in \set{\upair{1}{\infty},\upair{0}{\infty}} \union 
     \set{\upair{1}{1},\upair{1}{0},\upair{0}{1},\upair{0}{0}} \\ \\

\inference
{\Gamma= \hyp{x}{u}{\rgtype{r}{A}} \csum \Gamma' & \vlt{r}\\
R = \hyp{r}{\upair{v}{v'}}{A} \csum R' & v \neq 0, v'\neq \infty \\
R\Gives \Gamma& R';\Gamma'\Gives V:A}
{R;\Gamma \Gives \st{x}{V}:\tertype} 

\\ \\

\inference
{\Gamma= \hyp{x}{u}{\rgtype{r}{A}} \csum \Gamma' & \vlt{r}\\
R = \hyp{r}{\upair{v}{v'}}{A} \csum R' & v \neq 0, v'\neq \infty \\
R\Gives \Gamma & R';\Gamma'\Gives V:A}
{R;\Gamma \Gives \store{x}{V}:\behtype} 

\end{array}
\]%}
\caption{Restricted usages and rules for confluence}\label{rules-confluence}
\end{table}
We denote with $\CGives$ provability in this restricted system.
This system still enjoys the subject reduction property and moreover
its typable programs are strongly confluent.
\\
\begin{proposition}[subj. red. and confluence]\label{confluence-thm}
Suppose $R;\Gamma\CGives P:\alpha$. Then:\begin{enumerate}[(a)]
\item If $P\arrow P'$ then $R;\Gamma\CGives P':\alpha$
\item If $P\arrow P'$ and $P\arrow P''$ then either $P'\equiv P''$ or
there is a $Q$ such that $P'\arrow Q'$ and $P'' \arrow Q$
\end{enumerate}
\end{proposition}
\begin{proof}\ \\\vspace{-18pt}
  \begin{enumerate}[(a)]
  \item We just have to reconsider the case
where $E[\st{x}{V}] \arrow E[*] \mid \store{x}{V}$ and verify
that if $R;\Gamma \Gives \st{x}{V}:\tertype$ then 
$R;\Gamma \Gives \store{x}{V}:\behtype$ which entails that
$E[*]\mid \store{x}{V}$ is typable in the same context as 
$E[\st{x}{V}]$.
\item The restrictions on the usages forbid the typing of
a store such as the one in~\eqref{value-race1} and~\eqref{value-race2}  where two values are stored
in the same region. Moreover, it also forbids the typing of two parallel reads
on a volatile store~\eqref{volatile-store-race}.\end{enumerate}
%\vspace{-22pt}
\end{proof}

\begin{remark}
We note that the rules for ensuring confluence require that at most
one value is associated with a region (single-assignment). 
This is quite a restrictive discipline (comparable to the one in \cite{KPT99})
but one has to keep in mind that it targets regions that
can be accessed concurrently by several threads. Of course, the
discipline could be relaxed for the regions that are accessed by one
single sequential thread. Also, {\em e.g.}, for optimisation 
purposes, one may be interested in the confluence/determinism 
of certain reductions even when  the overall program 
is non-deterministic.  
\end{remark}

\section{An Affine-Intuitionistic Type and Effect System}
\label{aitype-effect-sec}
We refine the type system to include {\em effects} which are
denoted with $e,e',\ldots$ and are finite sets of regions. 
The syntax of programs (Table \ref{syntax-terms}) and their 
operational semantics (Table \ref{op-sem}) are unchanged.
The only modification to the syntax of types (Table \ref{syntax-types}) is that 
the affine implication is now annotated with an effect
so that we write:
$$A\limpe{e}\alpha$$
which is the type of a function that when given a value of type $A$
may produce something of type $\alpha$ and an effect on the regions in $e$.
This introduces a new dependency
of types on regions and consequently the compatibility condition between
region contexts and functional types in Table \ref{type-cxt-unstratified} 
becomes:
\[
\inference{R\dcl A\quad R\dcl \alpha \quad e\subseteq \w{dom}(R)}
{R\dcl (A\limpe{e} \alpha)}
\]
\begin{myexample}
One may verify that the judgement $$\hyp{r}{U}{\tertype \limpe{\set{r}} \tertype}\Gives $$ is 
derivable.  
\vspace{4pt}
\end{myexample}

The typing judgements now take the shape
$$R;\Gamma \Gives P:(\alpha,e)$$
where the effect $e$ provides an upper bound on the set of 
regions on which the program $P$ may read or write when it is
evaluated. In particular, we can be sure that values and stores
produce an empty effect. As for the operations to read and write
the store, one exploits the dependency of address types on regions
to determine the region where the effect occurs (cf. \cite{LG88}).
% % \begin{table}
% % %{\footnotesize
% % \[
% % \begin{array}{c c}
% % \inference{R;\Gamma,\hyp{x}{1}{A} \Gives M:(\alpha,e)}
% % {R;\Gamma \Gives \lambda x.M:( A \limpe{e} \alpha,\emptyset)}
% % &
% % \inference{
% % R_1;\Gamma_1\Gives M:(A\limpe{e} \alpha,e') \\
% % R_2;\Gamma_2\Gives N:(A,e'')
% % }
% % {R_1\csum R_2;\Gamma_1\csum \Gamma_2 \Gives MN:(\alpha,e\union
% %   e'\union e'')} \\ \\ 
% % \inference{
% % R\Gives \Gamma & \hyp{x}{u}{\rgtype{r}{A}}\in \Gamma  \\
% % \hyp{r}{\upair{v}{v'}}{A} \in R & v' \neq 0}
% % {R;\Gamma \Gives \get{x}:(A,\set{r})} 
% % &
% % \inference{
% % \Gamma= \hyp{x}{u}{\rgtype{r}{A}} \csum \Gamma'& \vlt{r} \\
% % R = \hyp{r}{\upair{v}{v'}}{A} \csum R' & v \neq 0 \\
% % R\Gives \Gamma& R';\Gamma'\Gives V:(A,\emptyset)}
% % {R;\Gamma \Gives \st{x}{V}:(\tertype,\set{r})} 
% % \end{array}
% % \]%}
% % \caption{Sample of the affine-intuitionistic type and effect system}\label{sample-ter-system}
% % \end{table}

The affine-intuitionistic type and effect system is spelled out in 
Table~\ref{ter-system}.
\begin{table}[t]
%{\footnotesize
\[
\begin{array}{c}

\inference{R\Gives \Gamma& \hyp{x}{u}{A}\in \Gamma}
{R;\Gamma \Gives x:(A,\emptyset)}

\qquad

\inference{R\Gives \Gamma}
{R;\Gamma \Gives *:(\tertype,\emptyset)} \\ \\

\inference{R;\Gamma,\hyp{x}{1}{A} \Gives M:(\alpha,e)}
{R;\Gamma \Gives \lambda x.M:( A \limpe{e} \alpha,\emptyset)}

\qquad

\inference{
R_1;\Gamma_1\Gives M:(A\limpe{e} \alpha,e') \\
R_2;\Gamma_2\Gives N:(A,e'')
}
{R_1\csum R_2;\Gamma_1\csum \Gamma_2 \Gives MN:(\alpha,e\union
  e'\union e'')} \\ \\ 

\inference{
R\csum R'\Gives (\Gamma\csum \Gamma')& \w{saff}(R';\Gamma')\\
R;\Gamma \Gives M:(A,e) & \neg \w{aff}(R;\Gamma)
%\hyp{x}{u}{B}\in \Gamma' \textrm{ implies }u=1
}
{R\csum R';\Gamma \csum \Gamma' \Gives \bang M:(\bang A,e)}

\qquad

\inference{
R_1;\Gamma_1\Gives M:(\bang A,e)\\
R_2;\Gamma_2,\hyp{x}{\infty}{A} \Gives (N,e'):\alpha
}
{R_1\csum R_2;\Gamma_1\csum \Gamma_2 \Gives
  \letm{x}{M}{N}:(\alpha,e\union e')} \\ \\

\inference{R;\Gamma,\hyp{x}{u}{\rgtype{r}{A}} \Gives P:(\alpha,e)}
{R;\Gamma \Gives \new{x}{P}:(\alpha,e)}

\qquad
\inference{
R\Gives \Gamma & \hyp{x}{u}{\rgtype{r}{A}}\in \Gamma  \\
\hyp{r}{\upair{v}{v'}}{A} \in R & v' \neq 0
}
{R;\Gamma \Gives \get{x}:(A,\set{r})} \\ \\

\inference{
\Gamma= \hyp{x}{u}{\rgtype{r}{A}} \csum \Gamma'& \vlt{r} \\
R = \hyp{r}{\upair{v}{v'}}{A} \csum R' & v \neq 0 \\
R\Gives \Gamma& R';\Gamma'\Gives V:(A,\emptyset)}
{R;\Gamma \Gives \st{x}{V}:(\tertype,\set{r})} 

\qquad

\inference{
\Gamma= \hyp{x}{u}{\rgtype{r}{\bang A}} \csum \Gamma' & \prs{r}\\
R = \hyp{r}{\upair{v}{v'}}{\bang A} \csum R' & v \neq 0 \\
R\Gives \Gamma& R';\Gamma'\Gives V:(\bang A,\emptyset)
}
{R;\Gamma \Gives \pst{x}{V}:(\tertype,\set{r})} \\ \\

\inference{
\Gamma= \hyp{x}{u}{\rgtype{r}{A}} \csum \Gamma' & \vlt{r}\\
R = \hyp{r}{\upair{v}{v'}}{A} \csum R' & v \neq 0 \\
R\Gives \Gamma& R';\Gamma'\Gives V:(A,\emptyset)
}
{R;\Gamma \Gives \store{x}{V}:(\behtype,\emptyset)} 

\qquad

\inference{
\Gamma= \hyp{x}{u}{\rgtype{r}{\bang A}} \csum \Gamma' & \prs{r}\\
R = \hyp{r}{\upair{v}{v'}}{\bang A} \csum R' & v \neq 0 \\
R\Gives \Gamma& R';\Gamma'\Gives V:(\bang A,\emptyset)
}
{R;\Gamma \Gives \pstore{x}{V}:(\behtype,\emptyset)} \\ \\

\inference{
R_1;\Gamma_1\Gives P:(\alpha,e)\\
R_2;\Gamma_2\Gives S:(\behtype,\emptyset)
}
{R_1\csum R_2; \Gamma_1\csum \Gamma_2 \Gives (P \mid S):(\alpha,e)}

\qquad

\inference{
R_i;\Gamma_i\Gives P_i:(\alpha_i,e_i)\\
P_i \textrm{ not a store } i=1,2
}
{R_1\csum R_2; \Gamma_1\csum \Gamma_2 \Gives (P_1 \mid
  P_2):(\behtype,e_1\union e_2)}

\end{array}
\]%}
\caption{An affine-intuitionistic type and effect system}\label{ter-system}
\end{table}
We stress that these rules are the same as the ones in Table
\ref{air-system} modulo the enriched syntax of the functional types and the
management of the effect $e$ on the right hand side of the 
sequents. The management of the effects is {\em additive} as in
\cite{LG88}, indeed effects are just {\em sets} of regions.

Also to allow for some flexibility, it is convenient to introduce a
subtyping relation on types and effects, that is to say on pairs $(\alpha,e)$, as specified in Table \ref{subtyping}.
\begin{table}[h]
%{\footnotesize
\[
\begin{array}{c c}

\inference{}
{R\Gives \alpha\leq \alpha}
&

\inference{R\Gives A\leq A'}
{R\Gives \bang A \leq \bang A'}

\\ \\

\multicolumn{2}{c}{\inference{
e\subseteq e'\subseteq\w{dom}(R)\\
R\Gives A'\leq A & R\Gives \alpha\leq \alpha'
}
{R\Gives (A\limpe{e}\alpha) \leq (A'\limpe{e'} \alpha')}}\\ \\ 

\inference{
e\subseteq e'\subseteq\w{dom}(R)\\
R\Gives \alpha\leq \alpha'
}
{R\Gives (\alpha,e)\leq (\alpha',e')} 

&

\inference{
R;\Gamma\Gives M:(\alpha,e)\\ 
R\Gives (\alpha,e)\leq (\alpha',e')
}
{R;\Gamma \Gives M:(\alpha',e')}

\end{array}
\]%}
\caption{Subtyping induced by effect containment}\label{subtyping}
\end{table}
We notice that the {\em transitivity rule} for subtyping
\[
\inference{R\Gives \alpha  \leq \alpha'\qquad R\Gives \alpha'\leq \alpha''}
{R\Gives \alpha \leq  \alpha''}
\]
can be derived via a simple induction on the height of the proofs.
\begin{remark}
The introduction of the subtyping rules has a limited impact
on the structure of the typing proofs. Indeed, if $R\Gives A\leq B$
then we know that $A$ and $B$ may just differ in the effects 
annotating the functional types. In particular, when looking at the
proof of the typing judgement of a value such as 
$R;\Gamma \Gives \lambda x.M: (A,e)$,  we can always argue that 
$A$ has the shape $A_1\limpe{e_{1}} A_2$ and, in case the effect $e$
is not empty, that there is a shorter proof of the judgement
$R;\Gamma \Gives \lambda x.M:(B_1\limpe{e_{2}} B_2,\emptyset)$ where
$R\Gives A_1\leq B_1$, $R\Gives B_2\leq A_2$, and $e_2\subseteq e_1$.  
\end{remark}

Then to prove subject reduction, we just repeat the proof of
Theorem \ref{sub-red-thm} while using standard arguments to keep 
track of the effects.
\begin{proposition}[subject reduction with effects]\label{sub-red-eff-thm}
Types and effects are preserved by reduction.
\end{proposition}
\begin{remark}
It is easy to check that a typable program such as $E[\st{x}{V}]$ which
is ready to produce an effect on the region $r$ associated with $x$
will indeed contain $r$ in its effect. Thus the subject reduction
property stated above as Proposition \ref{sub-red-eff-thm}
entails that the type and effect system does provide an upper bound on
the effects a program may produce during its evaluation.  
\end{remark}
%\ \\
\section{Termination}\label{termination-sec}
Terms typable in the unstratified type and effect system (cf. Table~\ref{ter-system})
may diverge, as exemplified here:
\begin{myexample}{}
\label{diverging-example}
The following term stores at the address $x$ a function 
that, given an argument, keeps fetching itself from the store forever:
\begin{equation*}
\new{x}{\pst{x}{\bang (\lambda y.\letm{x}{\get{x}}{xy  } )}  \ ; \ \letm{x}{\get{x}}{x*}}
\end{equation*}
One may verify that it is typable in a region context 
$$R=\hyp{r}{\upair{1}{\infty}}{\bang(\tertype \limpe{\set{r}} \tertype)}$$  
\end{myexample}

This example suggests that in order to recover termination,
we may order regions and make sure that
a value stored in a certain region when put in an
evaluation context can only produce effects on smaller regions.
This is where our type and effect system comes into play, and to formalise this idea, we introduce
in Table~\ref{type-cxt-stratified} rules for the formation of
types and contexts which are alternative to those in
Table~\ref{type-cxt-unstratified}.
\begin{myexample}{}
Assuming Table~\ref{type-cxt-stratified} and taking $R= \hyp{r}{U}{\tertype}$,
one may check that the judgement
$$\hyp{r}{U}{\tertype},\hyp{r'}{U'}{\tertype \limpe{\set{r}} \tertype}\Gives$$  is derivable
while $$\hyp{r'}{U'}{\tertype \limpe{\set{r'}}\tertype}\Gives$$ is
{\em not}. In particular, the region context of
Example~\ref{diverging-example} is neither derivable.
\vspace{4pt}  
\end{myexample}

\begin{table}[h]
%{\footnotesize
\[
\begin{array}{c c}

\inference{}{\emptyset \Gives}

&

\inference{R\Gives A& r\notin \w{dom}(R)}
{R,\hyp{r}{U}{A} \Gives} 

\\\\

\inference{R\Gives}
{R\Gives \tertype} 

&

\inference{R\Gives}
{R\Gives \behtype} 

\\\\
\inference{R\Gives A}
{R\Gives \bang A}

&

\inference{e\subseteq \w{dom}(R)\\
R\Gives A& R\Gives \alpha
}
{R\Gives (A\limpe{e} \alpha)} 

\\\\

\inference{R\Gives & \hyp{r}{U}{A} \in R}
{R\Gives \rgtype{r}{A}}  

&

\inference{R\Gives \alpha& e\subseteq\w{dom}(R)}
{R\Gives (\alpha,e)} 

\end{array}
\]%}
\caption{Formation of types and contexts (stratified)}\label{type-cxt-stratified}
\end{table}

It is easy to verify that the stratified system is a restriction of the
unstratified one and that the subject reduction (Proposition
\ref{sub-red-eff-thm}) still holds in the restricted stratified
system. If confluence is required, then one may add 
the restrictions spelled out in Table \ref{rules-confluence}.

Concerning termination, we recall that
there is a standard forgetful translation $(\_)$ from
affine-intuitionistic logic to intuitionistic logic which amounts to
forget about the modality $!$ and the usages and to regard the affine
implication as an ordinary intuitionistic implication.  Thus, for
instance, the translation of types goes as follows:
$\forget{\bang A} = \forget{A}$ and 
$\forget{A\limp B} = \forget{A}\arrow \forget{B}$;
while the translation of terms is:
$\forget{!M}=\forget{M}$ and  \quad 
$\forget{\letm{x}{M}{N}} = (\lambda x.\forget{N})\forget{M}$.
In Table \ref{forget-translation}, we lift this translation from the
stratified {\em affine-intuitionistic} type and effect system into a
stratified {\em intuitionistic} type and effect system defined in
\cite{Amadio09}.  
\begin{table}
%{\footnotesize
\[
\begin{array}{c}

\forget{\tertype} =\tertype,\quad
\forget{\behtype} = \behtype,\quad
\forget{A\limpe{e}\alpha} = \forget{A}\act{e} \forget{\alpha}, \quad
\forget{\bang A} = \forget{A},\quad 
\forget{\regtype{r}{A}} = \regtype{r}{\forget{A}} \\ \\

\forget{ \hyp{r_{1}}{U_{1}}{A_{1}},\ldots,\hyp{r_{n}}{U_{n}}{A_{n}}} = 
r_1:\forget{A_{1}},\ldots,r_n:\forget{A_{n}} \\ \\

\forget{\hyp{x}{u}{A},\Gamma} = \left\{
\begin{array}{ll} 
x:\forget{A},\forget{\Gamma} &\textrm{if }A\neq \regtype{r}{B} \\
\forget{\Gamma}              &\textrm{otherwise} 
\end{array}
\right. \\ \\

\forget{x} = x, 
\quad \forget{x^{r}} = r, 
\quad
\forget{*} = *, 
\quad
\forget{\lambda x.M} = \lambda x.\forget{M}, 
\quad
\forget{MN} = \forget{M}\forget{N} \\ \\

\forget{\bang M} = \forget{M},
\quad
\forget{\letm{x}{M}{N}} = (\lambda x.\forget{N})\forget{M}, 
\quad
\forget{\new{x}{M}} = \forget{M}, \\ \\

\forget{\get{x^{r}}} = \get{r}, 
\quad
\forget{\st{x^{r}}{V}} = \st{r}{\forget{V}},
\quad
\forget{\pst{x^{r}}{V}} = \pst{r}{\forget{V}}, \\ \\ 

\forget{\store{x^{r}}{V}} = \pstore{r}{\forget{V}},
\quad
\forget{\pstore{x^{r}}{V}} = \pstore{r}{\forget{V}},
\quad
\forget{P\mid P'} = \forget{P}\mid \forget{P'}

\end{array}
\]%}
\caption{Forgetful translation}\label{forget-translation}
\end{table}

The translation assumes a {\em decoration phase} where the (free or
bound) variables with a region type of the shape $\rgtype{r}{A}$ are
labelled with the region $r$. Intuitively, the intuitionistic system
abstracts an address $x$ related to the region $r$ to the
region $r$ itself so that a decorated variable $x^r$ translates into a
constant $r$. In the intuitionistic language, a region $r$ is a term of
region type $\rgtype{r}{A}$, for some $A$ and all stores are persistent.
The full definition of the language is recalled in
Appendix~\ref{termination-proof}.

It turns out that a reduction in the affine-intuitionistic system is
mapped to exactly a reduction in the intuitionistic system.  Then the
termination of the intuitionistic system proved in \cite{Amadio09}
entails the termination of the affine-intuitionistic one.

\begin{theorem}[termination]\label{termination-thm}
Programs typable in the stratified affine-intuitionistic 
type and effect system terminate.
\end{theorem}

%\ \\
\section{Conclusion}
We have presented an affine-intuitionistic system of types and effects for 
a functional-concurrent programming language. 
The main contribution over~\cite{Amadio09} is that the functional core
of the  system is based on Barber-Plotkin affine-intuitionistic logic
which distinguishes between affine and intuitionistic hypotheses. 
The `non-logical' part of the language, with operators
to read and write dynamically generated addresses of a `store', has
been refined to take into account the process of data duplication.
In the type system,  addresses are abstracted into a 
finite number of {\em regions}. We have introduced a 
suitable discipline of region {\em usage} and shown that it combines
with region {\em stratification} in the affine-intuitionistic setting
to regain  {\em confluence} and {\em termination}, respectively.

\paragraph{Future Work}
Beyond these crucial properties, we hope to show  
that other interesting properties of the functional core can be extended
to the considered framework. We think in particular of
the construction of denotational models (see, {\em e.g}, \cite{Bierman95})
and of bounds on the computational complexity of typable  programs 
(see, {\em e.g.}, \cite{Girard98}).

We also recall that more work would be required to get an operational programming
language, as with the introduction of inductive types and the
extension to a synchronous/timed framework (cf. \cite{Amadio09,Boudol07}) where determinism and
termination are useful properties.

%{\footnotesize
\section{Acknowledgements}
The first author was partially supported by ANR-06-SETI-010-02 and the second and
third authors by ANR-08-BLANC-0211-01.
%}

%{\footnotesize
%}

\newpage
\appendix

% \section{Affine-Intuitionistic type and effect system}\label{appendix-system}
% The typing rules of the affine-intuitionistic type and effect system 
% are spelled out in Table \ref{ter-system}. 

% \newpage

\section{Proofs}
\label{appendix-proof}

\subsection{Proof of Theorem \ref{sub-red-thm}}
\label{sub-red-proof}

\begin{lemma}[weakening]
  \label{weak-lem}
  If $R;\Gamma\vdash P:\alpha$ and $R\uplus R'\vdash \Gamma\uplus\Gamma'$ then $R\uplus R';\Gamma\uplus\Gamma'\vdash P:\alpha$.
\end{lemma}
\begin{proof}
  By induction on the typing of $P$. Following Table
\ref{air-system}, there are 14 rules to be considered of which 
we highlight 3.

\begin{description}

\item[$P \equiv MN$] We have:
    $$
    \inference
    {R_1;\Gamma_1\vdash M:A\multimap \alpha & R_2;\Gamma_2\vdash N:A}
    {R_1\uplus R_2;\Gamma_1\uplus\Gamma_2\vdash MN:\alpha} ~.
    $$
We notice that the composition operation $\uplus$ on contexts is
associative and commutative (when it is defined) and that 
$(R_1\uplus R_2 \uplus R')\Gives (\Gamma_1\uplus
\Gamma_2\uplus \Gamma')$ entails that 
$(R_1\uplus R') \Gives (\Gamma_1\uplus \Gamma')$. 
Hence, by induction hypothesis, we get 
    $R_1\uplus R';\Gamma_1\uplus\Gamma'\vdash M: A\multimap \alpha$, 
    from which we derive:
    $$
    \inference
    {R_1\uplus R';\Gamma_1\uplus\Gamma'\vdash M :A\multimap \alpha & 
     R_2;\Gamma_2\vdash N:A}
    {R_1\uplus R_2\uplus R';\Gamma_1\uplus\Gamma_2\uplus\Gamma'\vdash
    MN:\alpha} ~.
    $$

\item[$P \equiv \oc M$]  We have:
    $$
    \inference
    {R\uplus R''\vdash \Gamma\uplus\Gamma'' \\ \w{saff}(R'';\Gamma'')\\
      \neg \w{aff}(R;\Gamma) &       R;\Gamma\vdash M:A\\}
%      x:(u,B)\in\Delta \text{ implies } u=1}
    {R\uplus R'';\Gamma\uplus\Gamma''\vdash\oc M:\oc A}~.
    $$
    We can always decompose $R'$ as $R'_1\uplus R'_{\infty}$ and 
    $\Gamma'$ as $\Gamma'_1\uplus\Gamma'_{\infty}$ so that 
    $\neg \w{aff}(R'_{\infty};\Gamma'_{\infty})$ and 
     $\w{saff}(R'_1; \Gamma'_1)$.
    By induction hypothesis, we have $R\uplus
    R'_{\infty};\Gamma\uplus\Gamma'_{\infty}\vdash M:A$.
    We notice that $\neg \w{aff}(R\uplus
    R'_{\infty};\Gamma\uplus\Gamma'_{\infty})$ and 
    $\w{saff}(R'_1\uplus R'';\Gamma'_1\uplus \Gamma'')$ 
    (remember that $1\uplus \infty$ is undefined). Hence 
    we derive:
    $$
    \inference
    {(R\uplus R'_{\infty} \uplus R'_1\uplus R'') \Gives (\Gamma \uplus
      \Gamma'_{\infty} \uplus \Gamma'_1\uplus \Gamma'') \\ \w{saff}(R'_1\uplus R'';\Gamma'_1\uplus \Gamma'')\\
      \neg \w{aff}(R\uplus R'_{\infty};\Gamma\uplus\Gamma'_{\infty}) &
      R\uplus R'_{\infty};\Gamma\uplus\Gamma'_{\infty} \vdash M:A}
    {R\uplus R'\uplus R'';\Gamma\uplus\Gamma'\uplus\Gamma''\vdash
      \bang M:\bang A}~.
    $$

\item[$P \equiv \st{x}{V}$] We have:
    $$
    \inference
    {\Gamma=x:(u,\regtype{r}A)\uplus\Gamma''\\
      R=r:(\upair{v}{v'},A)\uplus R'' & v \neq 0\\
      R\vdash\Gamma & R'';\Gamma''\vdash V:A}
    {R;\Gamma \vdash \st{x}{V}:\tertype}~.
    $$
    By induction hypothesis, we have $R''\uplus R';\Gamma''\uplus\Gamma'\vdash V:A$, from which we derive:
    $$
    \inference
    {\Gamma\uplus \Gamma' =x:(u,\regtype{r}{A})\uplus(\Gamma''\uplus\Gamma')\\\
      R\uplus R' =r:(\upair{v}{v'},A)\uplus(R''\uplus R') & v \neq 0\\
      R\uplus R'\vdash\Gamma\uplus \Gamma' & 
     R''\uplus R';\Gamma''\uplus\Gamma'\vdash V:A}
    {R\uplus R';\Gamma\uplus \Gamma' \vdash \st{x}{V}:\tertype}~.
    $$
We notice that this argument still holds when introducing 
the restriction $v'\neq \infty$  in order to guarantee confluence
(cf. Table \ref{rules-confluence}). Indeed, the restriction 
$v'\neq \infty$ is equivalent to require that the
usage of the region $r$ ranges in the family of usages
$\set{\upair{1}{1},\upair{1}{0},\upair{0}{1},\upair{0}{0}}$.
  \end{description}
\end{proof}

\begin{lemma}[affine substitution lemma] \label{aff-sub-lem}
  If $R_1;\Gamma_1,x:(1,A) \vdash P:\alpha$,
     $R_2;\Gamma_2 \vdash V:A$,  and 
     $R_1 \uplus R_2 \vdash \Gamma_1 \uplus \Gamma_2$ then 
     $R_1 \uplus R_2;\Gamma_1 \uplus \Gamma_2 \vdash [V/x]P : \alpha$.
\end{lemma}
\begin{proof}
  By induction on the typing  of $P$. We highlight 
  4 cases out of 14.

  \begin{description}

  \item[$P \equiv MN$] We have:
    $$
    \inference
    {R_3;\Gamma'_3\vdash M:C\multimap \alpha & R_4;\Gamma'_4\vdash N:C}
    {R_3\uplus R_4;\Gamma'_3\uplus\Gamma'_4\vdash MN:\alpha}~.
    $$

Because $\hyp{x}{1}{A}$ is an affine hypothesis, it can occur 
exclusively either in $\Gamma'_3$ or in $\Gamma'_4$. We consider both cases.
    \begin{enumerate}

   \item $\Gamma'_3=\Gamma_3,x:(1,A)$ and $\Gamma'_4=\Gamma_4$ with $x
     \notin dom(\Gamma_4)$.
      By induction hypothesis we have $R_2\uplus R_3;\Gamma_2\uplus\Gamma_3\vdash [V/x]M:C\multimap \alpha$. Plus $x \notin \w{FV}(N)$ so $[V/x]N\equiv N$, hence $R_4;\Gamma_4\vdash [V/x]N:C$. Then we derive:
      $$
      \inference
      {R_2\uplus R_3;\Gamma_2\uplus\Gamma_3\vdash [V/x]M:C\multimap \alpha \\ R_4;\Gamma_4\vdash [V/x]N:C}
      {R_2\uplus R_3\uplus R_4;\Gamma_2\uplus\Gamma_3\uplus\Gamma_4\vdash [V/x](MN):\alpha}~.
      $$

\item $\Gamma'_3=\Gamma_3$ with $x \notin dom(\Gamma_3)$ and
  $\Gamma'_4=\Gamma_4,x:(1,A)$.

By induction hypothesis we have $R_2 \uplus
R_4;\Gamma_2\uplus\Gamma_4\vdash [V/x]N:C$. Plus $x \notin \w{FV}(M)$
so $[V/x]M\equiv M$, hence $R_3;\Gamma_3\vdash [V/x]M:C\multimap
\alpha$. Then we derive:
      $$
      \inference
      {R_3;\Gamma_3\vdash [V/x]M:C\multimap \alpha \\ R_2\uplus R_4;\Gamma_2\uplus\Gamma_4\vdash [V/x]N:C}
      {R_2\uplus R_3\uplus R_4;\Gamma_2\uplus\Gamma_3\uplus\Gamma_4\vdash [V/x](MN):\alpha}~.
      $$
    \end{enumerate}

    \item[$P\equiv\oc M$] We have:
      $$
      \inference{
          R_1\uplus R'\Gives (\Gamma_1\csum (\Gamma',x:(1,A))) \\ \w{saff}(R';\Gamma',x:(1,A))\\
          R_1;\Gamma_1 \Gives M:A & \neg \w{aff}(R_1;\Gamma_1)}
%          \hyp{x}{u}{B}\in \Delta'_1 \textrm{ implies }u=1
      {R_1\uplus R';\Gamma_1 \csum (\Gamma',x:(1,A)) \Gives \bang M:\bang A}
      $$
      We deduce that $x \notin \w{FV}(\oc M)$, hence $[V/x](\oc M)
      \equiv \oc M$ and $R_1\uplus R';\Gamma_1\csum\Gamma'\vdash
      [V/x](\oc M):\oc A$. By lemma \ref{weak-lem}, we get $R_1\uplus
      R'\uplus R_2;\Gamma_1\uplus\Gamma'\csum\Gamma_2\vdash [V/x](\oc
      M):\oc A$. \\

  \item[$P \equiv \letb{y}{M}{N}$] Renaming $y$ so that $y\neq x$, we have:
    $$
    \inference
    {R_3;\Gamma'_3\vdash M:\oc C & R_4;\Gamma'_4,y:(\infty,C)\vdash N:\alpha}
    {R_3\uplus R_4;\Gamma'_3\uplus\Gamma'_4\vdash \letb{y}{M}{N}:\alpha}
    $$
    As in the case of application, we distinguish two cases.

    \begin{enumerate}

    \item $\Gamma'_3=\Gamma_3,x:(1,A)$ and $\Gamma'_4=\Gamma_4$ with $x \notin dom(\Gamma_4)$.\\
      By induction hypothesis, we have $R_2\uplus R_3;\Gamma_2\uplus\Gamma_3\vdash [V/x]M:\oc C$. Plus $x \notin \w{FV}(N)$ so $[V/x]N\equiv N$, hence $R_4;\Gamma_4,y:(\infty,C)\vdash [V/x]N:\alpha$. Then we derive:
      $$
      \inference
      {R_2\uplus R_3;\Gamma_2\uplus\Gamma_3\vdash [V/x]M:\oc C \\ R_4;\Gamma_4,y:(\infty,C)\vdash [V/x]N:\alpha}
      {R;\Gamma_2\uplus\Gamma_3\uplus\Gamma_4\vdash [V/x](\letb{y}{M}{N}):\alpha}~.
      $$
      where $R=R_2\uplus R_3\uplus R_4$.
    \item $\Gamma'_3=\Gamma_3$ with $x \notin dom(\Gamma_3)$ and $\Gamma'_4=\Gamma_4,x:(1,A)$.\\
      By induction hypothesis we have $R_2\uplus
      R_4;\Gamma_2,y:(\infty,C)\uplus\Gamma_4\vdash
      [V/x]N:\alpha$. Plus $x \notin \w{FV}(M)$ so $[V/x]M \equiv M$, hence $R_3;\Gamma_3\vdash [V/x]M: \oc C$. Then we derive:
      $$
      \inference
      {R_3;\Gamma_3\vdash [V/x]M: \oc C \\ R_2\uplus R_4;\Gamma_2,y:(\infty,C)\uplus\Gamma_4\vdash [V/x]N:\alpha}
      {R;\Gamma_2\uplus\Gamma_3\uplus\Gamma_4\vdash [V/x](\letb{y}{M}{N}):\alpha}~.
      $$
      where $R=R_2\uplus R_3\uplus R_4$.
    \end{enumerate}

  \item[$P \equiv \st{y}{V'}$] We distinguish two cases. \\

\begin{enumerate}

\item  If $y \neq x$ we have:
    $$
    \inference
    {\Gamma_1,x:(1,A)=y:(u,\regtype{r}C)\uplus\Gamma'_1\\
      R_1=r:(\upair{v}{v'},C)\uplus R'_1 & v \neq 0\\
      R_1\vdash\Gamma_1,x:(1,A) & R'_1;\Gamma'_1\vdash V':C}
    {R_1;\Gamma_1,x:(1,A) \vdash \st{y}{V'}:\tertype}~.
    $$
    We deduce that $\Gamma'_1=\Gamma''_1\uplus x:(1,A)$, and by
    induction hypothesis we get $R'_1\uplus
    R_2;\Gamma''_1\uplus\Gamma_2\vdash [V/x]V' :C$, from which we
    derive:
    $$
    \inference
    {\Gamma_1=y:(u,\regtype{r}C)\uplus\Gamma''_1\\
      R_1=r:(\upair{v}{v'},C)\uplus R'_1 & v \neq 0\\
      R_1\vdash\Gamma_1 & R'_1\uplus R_2;\Gamma''_1\uplus\Gamma_2\vdash [V/x]V':C}
    {R_1;\Gamma_1 \vdash [V/x]\st{y}{V'}:\tertype}~.
    $$
    By lemma \ref{weak-lem}, 
we obtain $$R_1\uplus R_2;\Gamma_1\uplus\Gamma_2 \vdash [V/x]\st{y}{V'}:\tertype$$.
    
\item    If $y = x$ then $[V/x]\st{y}{V'} \equiv \st{V}{V'}$,
    $A=\regtype{r}C$, and $u=1$. Moreover $V$ must be a variable, 
thus we can derive:
    $$
    \inference
    {\Gamma_1=V:(1,\regtype{r}C)\uplus\Gamma'_1\\
      R_1=r:(\upair{v}{v'},C)\uplus R'_1 & v \neq 0\\
      R_1\vdash\Gamma_1 & R'_1;\Gamma'_1\vdash V':C}
    {R_1;\Gamma_1 \vdash [V/x]\st{y}{V'}:\tertype} ~,
    $$
    and by lemma \ref{weak-lem} we get $$R_1\uplus R_2;\Gamma_1\uplus\Gamma_2 \vdash [V/x]\st{y}{V'}:\tertype$$.

\end{enumerate}
  \end{description}
\end{proof}

\begin{lemma}[intuitionistic substitution lemma]
\label{int-sub-lem}
  If \\$R_1;\Gamma_1,x:(\infty,A) \vdash P:\alpha$, 
     $R_2;\Gamma_2 \vdash \oc V:\oc A$, and 
     $R_1 \uplus R_2 \vdash \Gamma_1 \uplus \Gamma_2$ 
  then $R_1 \uplus R_2;\Gamma_1 \uplus \Gamma_2 \vdash [V/x]P : \alpha$.
\end{lemma}
\begin{proof}
  By induction on the typing of $P$. We highlight 4 cases out of 14.

  \begin{description}

  \item[$P \equiv MN$] We have:
    $$
    \inference
    {R_3;\Gamma'_3\vdash M:C\multimap \alpha & R_4;\Gamma'_4\vdash N:C}
    {R_3\uplus R_4;\Gamma'_3\uplus\Gamma'_4\vdash MN:\alpha}~.
    $$
    We distinguish 3 cases.
    \begin{enumerate}

    \item 
$\Gamma'_3=\Gamma_3,x:(\infty,A)$ and 
$\Gamma'_4=\Gamma_4$ with $x \notin dom(\Gamma_4)$.\\
     By induction hypothesis we have $R_2\uplus
     R_3;\Gamma_2\uplus\Gamma_3\vdash [V/x]M:C\multimap \alpha$. Plus
     $x \notin \w{FV}(N)$ so $[V/x]N\equiv N$, hence
     $R_4;\Gamma_4\vdash [V/x]N:C$. Then we derive:
      $$
      \inference
      {R_2\uplus R_3;\Gamma_2\uplus\Gamma_3\vdash [V/x]M:
       C\multimap \alpha \\ R_4;\Gamma_4\vdash [V/x]N:C}
      {R_2\uplus R_3\uplus R_4;\Gamma_2\uplus\Gamma_3\uplus\Gamma_4
       \vdash [V/x](MN):\alpha}~.
      $$

    \item 
$\Gamma'_3=\Gamma_3$ with $x \notin dom(\Gamma_3)$ and 
$\Gamma'_4=\Gamma_4,x:(\infty,A)$.\\
      By induction hypothesis we have $R_2\uplus
      R_4;\Gamma_2\uplus\Gamma_4\vdash [V/x]N:C$. Plus $x \notin
      \w{FV}(M)$ so $[V/x]M\equiv M$, hence $R_3;\Gamma_3\vdash
      [V/x]M:C\multimap \alpha$. Then we derive:
      $$
      \inference
      {R_3;\Gamma_3\vdash [V/x]M:C\multimap \alpha \\ R_2\uplus R_4;\Gamma_2\uplus\Gamma_4\vdash [V/x]N:C}
      {R_2\uplus R_3\uplus R_4;\Gamma_2\uplus\Gamma_3\uplus\Gamma_4\vdash [V/x](MN):\alpha}~.
      $$
    \item 
$\Gamma'_3 = \Gamma_3,x:(\infty,A)$ and 
$\Gamma'_4 = \Gamma_4,x:(\infty,A)$.\\ 
By induction hypothesis we have
      $R_2\uplus R_3;\Gamma_2\uplus\Gamma_3\vdash [V/x]M:C\multimap
      \alpha$ and $R_2\uplus R_4;\Gamma_2\uplus\Gamma_4 \vdash
      [V/x]N:C$. Moreover we have:
      $$
      \inference
      {R_5\uplus R'\vdash \Gamma_5\uplus\Gamma' & \w{saff}(R';\Gamma')\\
        R_5;\Gamma_5\vdash V:A & \neg \w{aff}(R_5;\Gamma_5)}
%        x:(u,C) \in \Delta' \text{ implies } u=1}
      {R_2;\Gamma_2\vdash\oc V:\oc A}~,
      $$
      where $R_2=R_5\uplus R'$ and $\Gamma_2=\Gamma_5\uplus\Gamma'$.
Hence we know that all the hypotheses of $R'$ and $\Gamma'$ are of
weakened regions and variables. Thus we also have $R_3\uplus
R_5;\Gamma_3\uplus\Gamma_5\vdash [V/x]M:C\multimap \alpha$ and
$R_4\uplus R_5;\Gamma_4\uplus\Gamma_5\vdash [V/x]N:C$. Plus from $\neg
\w{aff}(R_5;\Gamma_5)$ we get $R_5\uplus R_5=R_5$ and
$\Gamma_5\uplus\Gamma_5=\Gamma_5$, and we can derive:
      $$
      \inference
      {R_3\uplus R_5;\Gamma_3\uplus\Gamma_5\vdash [V/x]M:C\multimap \alpha \\ R_4\uplus R_5;\Gamma_4\uplus\Gamma_5\vdash [V/x]N:C}
      {R_3\uplus R_4\uplus R_5;\Gamma_3\uplus\Gamma_4\uplus\Gamma_5\vdash [V/x](MN):\alpha}~.
      $$
      By lemma \ref{weak-lem} we obtain $R_2\uplus R_3\uplus R_4;\Gamma_2\uplus\Gamma_3\uplus\Gamma_4\vdash [V/x](MN):\alpha$.
    \end{enumerate}

\item[$P\equiv\oc M$] Suppose:
        $$
        \inference{
          R_5\uplus R'\Gives (\Gamma_5,x:(\infty,A))\uplus\Gamma'
          & \w{saff}(R';\Gamma')\\
          R_5;\Gamma_5,x:(\infty,A) \Gives M:B \\ \neg \w{aff}(R_5;\Gamma_5,x:(\infty,A))
%          \hyp{x}{u}{C}\in \Delta'_1 \textrm{ implies }u=1
        }
      {R_5\uplus R';(\Gamma_5,x:(\infty,A))\uplus\Gamma' \Gives \bang M:\bang B}~.
      $$
      And also:
      $$
      \inference
      {R_6\uplus R_7\vdash\Gamma_6\uplus\Gamma_7 &
        \w{saff}(R_7;\Gamma_7)\\
        \w{aff}(R_6;\Gamma_6) & R_6;\Gamma_6\vdash V:A}
      {R_2;\Gamma_2\vdash \bang V:\bang A}~,
      $$
      with $R_2=R_6\uplus R_7$ and
      $\Gamma_2=\Gamma_6\uplus\Gamma_7$. Hence we know that all the
      hypotheses of $R_7$ and $\Gamma_7$ are of weakened regions and
      variables, such that $R_6;\Gamma_6\vdash\bang V:\bang A$. By induction hypothesis we get $R_5\uplus R_6;\Gamma_5\uplus\Gamma_6 \vdash [V/x]M:B$ and we can derive:
        $$
        \inference
        {(R_5 \uplus R_6)\uplus (R_7\uplus R')\vdash (\Gamma_5 \uplus
          \Gamma_6)\uplus (\Gamma_7\uplus \Gamma') \\
          \w{saff}(R_7\uplus R';\Gamma_7\uplus\Gamma')\\
        \neg \w{aff}( R_5\uplus R_6;\Gamma_5\uplus\Gamma_6) \\  R_5\uplus R_6;\Gamma_5\uplus\Gamma_6 \vdash [V/x]M:B }
        {R_5\uplus R_2\uplus
        R';\Gamma_5\uplus\Gamma_2\uplus\Gamma'\vdash
[V/x]\oc M:\oc B}~.
          $$

  \item[$P \equiv \letb{y}{M}{N}$] We have:
    $$
    \inference
    {R_3;\Gamma'_3\vdash M:\oc C & R_4;\Gamma'_4,y:(\infty,C)\vdash N:\alpha}
    {R_3\uplus R_4;\Gamma'_3\uplus\Gamma'_4\vdash \letb{y}{M}{N}:\alpha}~.
    $$
    with $y\neq x$. We just spell out the case where $\Gamma'_3 =
    \Gamma_3,x:(\infty,A)$ and
    $\Gamma'_4 =\Gamma_4,x:(\infty,A)$.
By induction hypothesis, we have $R_2\uplus
R_3;\Gamma_2\uplus\Gamma_3\vdash [V/x]M:\oc C$ and $R_2\uplus
R_4;(\Gamma_2,y:(\infty,C))\uplus\Gamma_4\vdash
[V/x]N:\alpha$. Moreover we have:
      $$
      \inference
      {R_5\uplus R'\vdash \Gamma_5\uplus\Gamma'\quad \w{saff}(R';\Gamma')\\
        R_5;\Gamma_5\vdash V:A & \neg \w{aff}(R_5;\Gamma_5)}
      {R_2;\Gamma_2\vdash\oc V:\oc A}~,
      $$ 
where $\Gamma_2=\Gamma_5\uplus\Gamma'$ and $R_2=R_5\uplus
      R'$. Hence we know that all the hypotheses of $R'$ and $\Gamma'$
      are of weakened regions and variables. Thus we also have
      $R_3\uplus R_5;\Gamma_3\uplus\Gamma_5\vdash [V/x]M:\bang C$ and
      $R_4\uplus R_5;(\Gamma_4,y:(\infty,C))\uplus\Gamma_5\vdash
      [V/x]N:\alpha$. Plus from $\neg \w{aff}(R_5;\Gamma_5)$ we get
      $\Gamma_5\uplus\Gamma_5=\Gamma_5$ and $R_5\uplus R_5=R_5$, and
      we can derive:
      $$
\infer{
\begin{array}{c}
R_3\uplus R_5;\Gamma_3\uplus\Gamma_5\vdash [V/x]M:\bang C \\
R_4\uplus R_5;(\Gamma_4,y:(\infty,C))\uplus\Gamma_5\vdash
      [V/x]N:\alpha
\end{array}}
{R_3\uplus R_4\uplus R_5;\Gamma_3\uplus\Gamma_4\uplus\Gamma_5\vdash [V/x](\letb{y}{M}{N}):\alpha}~.
      $$
By lemma \ref{weak-lem}, we obtain $R_2\uplus R_3\uplus
R_4;\Gamma_2\uplus\Gamma_3\uplus\Gamma_4\vdash
[V/x](\letb{y}{M}{N}):\alpha$.

 \item[$P \equiv \st{y}{V'}$] We just look at the case $y\neq x$. We
    have:
    $$
    \inference
    {\Gamma_1,x:(\infty,A)=y:(u,\regtype{r}C)\uplus\Gamma'_1\\
      R_1=r:(\upair{v}{v'},C)\uplus R'_1 & v' \neq 0\\
      R_1\vdash\Gamma_1,x:(\infty,A) & R'_1;\Gamma'_1\vdash V':C}
    {R_1;\Gamma_1,x:(\infty,A) \vdash \st{y}{V'}:\tertype}~.
    $$
    We deduce that $\Gamma'_1=\Gamma''_1\uplus x:(\infty,A)$, and by
    induction hypothesis we get $R'_1\uplus
    R_2;\Gamma''_1\uplus\Gamma_2\vdash [V/x]V':C$, from which we
    derive:
    $$
    \inference
    {\Gamma_1=y:(u,\regtype{r}C)\uplus\Gamma''_1\\
      R_1=r:(\upair{v}{v'},C)\uplus R'_1 & v' \neq 0\\
      R_1\vdash\Gamma_1 & R'_1\uplus R_2;\Gamma''_1\uplus\Gamma_2\vdash [V/x]V':C}
    {R_1 \uplus R_2 ;\Gamma_1\uplus \Gamma_2 \vdash [V/x]\st{y}{V'}:\tertype}    ~.
    $$

\end{description}
\end{proof}

\begin{lemma}[structural equivalence preserves typing] \label{sub-red-equ}
If $R;\Gamma\vdash P:\alpha$ and $P\equiv P'$ then $R;\Gamma\vdash P':\alpha$.
\end{lemma}
\begin{proof}
Recall that structural equivalence is the least equivalence
relation induced by the equations stated in 
Table \ref{op-sem} and closed under static contexts.
Then we proceed by induction on the proof of structural equivalence.
This is is mainly a matter of reordering the pieces of the typing
proof of $P$ so as to obtain a typing proof of $P'$.
\end{proof}

\begin{lemma}[evaluation contexts and typing]  \label{eva-sub-lem}
Suppose that in the proof of $R;\Gamma \Gives E[M]:\alpha$ 
we prove $R';\Gamma' \Gives M:A$. Then replacing $M$ with a 
$M'$ such that $R';\Gamma'\Gives M':A$, we can still derive
$R;\Gamma \Gives E[M']:\alpha$.
\end{lemma}
\begin{proof}
By induction on the structure of $E$.
\end{proof}

\begin{lemma}[functional redexes]\label{fun-redex}
If $R;\Gamma \Gives E[\Delta] :\alpha$ where 
$\Delta$ has the shape $(\lambda x.M)V$ or $\letm{x}{V}{M}$ then
$R;\Gamma \Gives E[[V/x]M]:\alpha$.
\end{lemma}
\begin{proof}
If $\Delta = (\lambda x.M)V$ we appeal to the affine substitution
lemma \ref{aff-sub-lem} and if $\Delta=\letm{x}{V}{M}$ we rely on the
intuitionistic lemma \ref{int-sub-lem}. This settles the case
where the evaluation context $E$ is trivial. If it is complex
then we also need  lemma \ref{eva-sub-lem}.
\end{proof}

\begin{lemma}[side-effects redexes]\label{side-eff-redex}
If $R;\Gamma \Gives \Delta:\alpha$ where
$\Delta$ is one of the programs on the left-hand
side then $R;\Gamma \Gives \Delta':\alpha$ where
$\Delta'$ is the corresponding program on the right-hand side:
\[
\begin{array}{lc|c}

(1) &E[\st{x}{V}]       &E[*]\mid \store{x}{V} \\
(2) &E[\pst{x}{V}]       &E[*]\mid \pstore{x}{V} \\
(3) &E[\get{x}] \mid \store{x}{V}  &E[V] \\
(4)\quad &E[\get{x}] \mid \pstore{x}{\bang V}  \quad & \quad E[\bang V] \mid \pstore{x}{\bang V}

\end{array}
\]
\end{lemma}
\begin{proof}
We proceed by case analysis.

\begin{enumerate}

\item Suppose we derive $R;\Gamma \Gives E[\st{x}{V}]:\alpha$ from
$R_2;\Gamma_2 \Gives \st{x}{V}:\tertype$.
By the typing rule for $\st{x}{V}$ we know that 
$R_2= \hyp{r}{\upair{v}{v'}}{A} \csum R_3$, $\vlt{r}$,
$\Gamma_2 = \hyp{x}{u}{\rgtype{r}{A}} \csum \Gamma_3$, and
$R_3;\Gamma_3 \Gives V:A$.
It follows that $R_2;\Gamma_2 \Gives \store{x}{V}:\behtype$.
We can decompose $R_2;\Gamma_2$ into an additive part 
$(R_{2};\Gamma_2)^{0}$ and a multiplicative one 
$(R_{2};\Gamma_2)^{1}$.
Then from $(R_{2};\Gamma_{2})^{0} \Gives *:\tertype$,
we can derive $R_1;\Gamma_1\Gives E[*]:\alpha$, where
$(R_1;\Gamma_1) \csum (R_2;\Gamma_2)^{1} = R;\Gamma$.

\item Suppose we derive $R;\Gamma \Gives E[\pst{x}{V}]:\alpha$ from
$R_2;\Gamma_2 \Gives \pst{x}{V}:\tertype$.
By the typing rule for $\pst{x}{V}$ we know that 
$R_2= \hyp{r}{\upair{v}{v'}}{\bang A} \csum R_3$, $\prs{r}$,
$\Gamma_2 = \hyp{x}{u}{\rgtype{r}{\bang A}} \csum \Gamma_3$, and
$R_3;\Gamma_3 \Gives V:\bang A$.
It follows that $R_2;\Gamma_2 \Gives \pstore{x}{V}:\behtype$.
Then we reason as in the previous case.

\item Suppose $R_1;\Gamma_1 \Gives E[\get{x}]:\alpha$ is derived from
$R_2;\Gamma_2 \Gives \get{x}:A$, that $R_3;\Gamma_3 \Gives \store{x}{V}:\behtype$,
and that $R;\Gamma = (R_1;\Gamma_1) \csum (R_3;\Gamma_3)$.
Then $(R_2;\Gamma_2) \csum (R_3;\Gamma_3) \Gives V: A$, by weakening.
Also, let $r$ be the region associated with the address $x$. We know
that $\vlt{r}$ and that $R_2$ must have a reading usage on $r$.
It follows that $\w{aff}(R_2;\Gamma_2)$ and therefore the context
$E$ {\em cannot} contain a $\bang$. 
Thus from   $(R_2;\Gamma_2) \csum (R_3;\Gamma_3) \Gives V: A$ we
can derive $R;\Gamma \Gives E[V]:\alpha$.

\item  Suppose $R_1;\Gamma_1 \Gives E[\get{x}]:\alpha$ is derived from
$R_2;\Gamma_2 \Gives \get{x}:\bang A$, 
that $R_3;\Gamma_3 \Gives \pstore{x}{\bang V}:\behtype$,
and that $R;\Gamma = (R_1;\Gamma_1) \csum (R_3;\Gamma_3)$.
By the promotion rule, $R_3;\Gamma_3$ is a weakening of 
$R_4;\Gamma_4$ such that $\neg \w{aff}(R_4;\Gamma_4)$ and 
$R_4;\Gamma_4\Gives V: A$. 
Then from $R_4;\Gamma_4 \Gives \bang V:\bang A$ we can derive
$R';\Gamma' \Gives E[\bang V]:\alpha$ where $R;\Gamma$ is a
weakening of 
$(R';\Gamma')\csum (R_3;\Gamma_3)$.

\end{enumerate}
\end{proof}

\begin{theorem}[subject reduction]
  If $R;\Gamma\vdash P:\alpha$ and  $P \rightarrow P'$ then 
  $R;\Gamma\vdash P':\alpha$.
\end{theorem}
\begin{proof}
We recall that $P\arrow P'$ means that 
$P$ is structurally equivalent to a program
$C[\Delta]$ where $C$ is a static context, $\Delta$ 
is one of the programs on the left-hand side of the
rewriting rules specified in Table \ref{op-sem},
$\Delta'$ is the respective program on the right-hand side,
and $P'$ is syntactically equal to $C[\Delta']$.

By lemma \ref{sub-red-equ}, we know that 
$R;\Gamma \Gives C[\Delta]:\alpha$.
This entails that $R';\Gamma'\Gives \Delta : \alpha'$ for
suitable $R',\Gamma',\alpha'$.
By lemmas \ref{fun-redex} and \ref{side-eff-redex}, we derive that
$R';\Gamma'\Gives \Delta':\alpha'$.
Then by induction on the structure of $C$ we argue 
that $R;\Gamma \Gives C[\Delta']:\alpha$. 
\end{proof}

\subsection{Proof of Theorem \ref{termination-thm}}
\label{termination-proof}

Table \ref{summary1} summarizes the main syntactic categories and
the reduction rules of the intuitionistic system. 
It is important to notice that in the intuitionistic system
regions are terms and that the operations that manipulate the store
operate directly on the regions so that we write:
$\get{r}$, $\pst{r}{V}$, and $\pstore{r}{V}$ rather than
$\get{x}$, $\pst{x}{V}$, and $\pstore{x}{V}$. 

\begin{table}[h]
%{\footnotesize
\begin{center}
{\sc Syntax: terms}
\end{center}
\[
\begin{array}{ll}

x,y,\ldots         
&\textrm{(Variables)} \\

r,s,\ldots         
&\textrm{(Regions)} \\

V::= x \Alt * \Alt r \Alt \lambda x.M
&\textrm{(Values)} \\

M::= V \Alt MM \Alt \get{V} \Alt \pst{V}{V} 
\Alt (M \mid M)\quad &\textrm{(Terms)} \\

S::= \pstore{r}{v} \Alt (S \mid S) &\textrm{(Stores)} \\

P::= M \Alt S \Alt (P\mid P) &\textrm{(Programs)}  \\

E::= [~] \Alt EM \Alt VE  &\textrm{(Evaluation Contexts)} \\

C::= [~] \Alt (C \mid P)  (P\mid C) &\textrm{(Static Contexts)}   

\end{array}
\]
\begin{center}
{\sc Operational semantics}
\end{center}
\[
\begin{array}{cccc}

P\mid P' &\equiv  &P' \mid P    &\textrm{(Commutativity)} \\
(P\mid P')\mid P'' &\equiv &P\mid (P' \mid P'')  \quad &\textrm{(Associativity)} 

\end{array}
\]
\[
\begin{array}{ccc}

E[(\lambda x.M)V]        &\arrow  &E[[V/x]M]  \\

E[\get{r}],\pstore{r}{V} &\arrow  &E[V],\pstore{r}{V} \\ 

E[\pst{r}{V}]            &\arrow  &E[*],\pstore{r}{V}

\end{array}
\]
\begin{center}
{\sc Syntax: types and contexts}
\end{center}
\[
\begin{array}{ll}

\alpha::= A \Alt \behtype       &\textrm{(Types)} \\

A::= \tertype \Alt  (A \act{e} \alpha) \Alt \regtype{r}{A} 
&\textrm{(Value-types)} \\

\Gamma::= x_1:A_1,\ldots,x_n:A_n
&\textrm{(Contexts)} \\

R::= r_1:A_1,\ldots,r_n:A_n \quad &\textrm{(Region contexts)} 

\end{array}
\]%}
\caption{Intuitionistic system: syntactic categories and operational semantics}\label{summary1}
\end{table}

Table \ref{summary2} summarizes the typing rules for the 
stratified type and effect system.

\paragraph{Proviso}
To avoid confusion, in the following we write $\AIGives$ for provability 
in the affine-intuitionistic system and $\IGives$ for provability
in the intuitionistic system. \\

\begin{table}[h]
%{\footnotesize
\begin{center}
\mbox{{\sc Stratified region contexts and types}}  \\
\end{center}
\[
\begin{array}{c}

\infer{}{\emptyset \Gives}

\qquad

\infer{R\Gives A\quad r\notin \w{dom}(R)}
{R,r:A\Gives} 
\qquad

\infer{R\Gives}
{R\Gives \tertype} 

\qquad

\infer{R\Gives}
{R\Gives \behtype}
\\ \\ 

\infer{R\Gives A\quad R\Gives \alpha \quad e\subseteq \w{dom}(R)}
{R\Gives (A\act{e} \alpha)} 

\qquad

\infer{R\Gives \quad r:A \in R}
{R\Gives \regtype{r}{A}} 

\qquad

\infer{R\Gives \alpha \quad e\subseteq\w{dom}(R)}
{R\Gives (\alpha,e)} \\  \\

\textrm{{\sc Subtyping rules}} \\

\infer{R\Gives \alpha}{R\Gives \alpha \leq \alpha}

\qquad

\infer{\begin{array}{c}
R\Gives A'\leq A\quad  R\Gives \alpha \leq \alpha'\\
e\subseteq e' \subseteq \w{dom}(R) 
\end{array}}
{R\Gives (A\act{e} \alpha) \leq (A' \act{e'} \alpha')} 

\\ \\

\infer{\begin{array}{c}
R\Gives \alpha \leq \alpha'\\ 
e\subseteq e' \subseteq \w{dom}(R)
\end{array}}
{R\Gives (\alpha,e) \leq (\alpha',e')} 

\qquad

\infer{R;\Gamma \Gives M:(\alpha,e) \quad R\Gives (\alpha,e)\leq (\alpha',e')}
{R;\Gamma \Gives M:(\alpha',e')} 

\\ \\

\textrm{{\sc Terms, stores, and programs}} \\ 

\infer{R\Gives \Gamma \quad x:A\in \Gamma}
{R;\Gamma\Gives x:(A,\emptyset)}

\qquad 
\infer{R\Gives \Gamma \quad r:A\in R}
{R;\Gamma\Gives r:(\regtype{r}{A},\emptyset)} 

\qquad
\infer{R\Gives \Gamma}
{R;\Gamma \Gives *: (\tertype,\emptyset)} \\ \\

\infer{R;\Gamma,x:A \Gives M: (\alpha,e)}
{R;\Gamma\Gives \lambda x.M :(A\act{e} \alpha,\emptyset)}

\qquad

\infer{R;\Gamma \Gives M: (A\act{e_{2}} \alpha ,e_{1}) \quad R;\Gamma\Gives N: (A,e_{3})}
{R;\Gamma \Gives MN :(\alpha,e_1\union e_2\union e_3)} \\ \\

\infer{R;\Gamma \Gives V: (\regtype{r}{A}, \emptyset)}
{R;\Gamma \Gives \get{V} : (A,\set{r})} 

\qquad

\infer{R;\Gamma \Gives V: (\regtype{r}{A}, \emptyset)\quad R;\Gamma \Gives V':(A,\emptyset)}
{R;\Gamma \Gives \pst{V}{V'} : (\tertype, \set{r})} \\ \\

\infer{r:A \in R \quad R;\Gamma \Gives V: (A,\emptyset)}
{R;\Gamma \Gives \pstore{r}{V} :(\behtype,\emptyset)}

\qquad 

\infer{\begin{array}{c}
R;\Gamma \Gives P:(\alpha,e) \\
R;\Gamma \Gives S:(\behtype,\emptyset)
\end{array}}
{R;\Gamma \Gives (P\mid S):(\alpha, e)}

\\ \\
\infer{
P_i \textrm{ not a store} \quad 
R;\Gamma \Gives P_i:(\alpha_i,e_i), \quad i=1,2
}
{R;\Gamma \Gives (P_1\mid P_2):(\behtype,e_1\union e_2)}

\end{array}
\]%}
\caption{Intuitionistic system: stratified types and effects}\label{summary2}
\end{table}

The translation acts on typable programs.  In order to define it, it
is useful to go through a phase of {\em decoration} which amounts to label
each occurrence (either free or bound) of a variable $x$ of region 
type $\regtype{r}{A}$ with the region $r$.  
For instance,  suppose $R=\hyp{r_{1}}{U_{1}}{A_{1}},\ldots,\hyp{r_{4}}{U_{4}}{A_{4}}$ and
suppose we have a provable judgement:
\[
\begin{array}{l}
R; \hyp{x_{1}}{u_{1}}{\regtype{r_{1}}{A}} \AIGives \\
\qquad x_1 \mid \letm{x_{2}}{\ldots}{x_{2}} \mid 
\lambda x_3.x_3 \mid \new{x_{4}}{x_4} : (\behtype,\emptyset)  
\end{array}
\]
Further suppose in the proof the variable $x_i$ relates to the region $r_i$ for $i=1,\ldots,4$.
Then the decorated term is:
\[
x_{1}^{r_{1}} \mid \letm{x_{2}}{\ldots}{x_{2}^{r_{2}}} \mid 
\lambda x_3.x_{3}^{r_{3}} \mid \new{x_{4}}{x_{4}^{r_{4}}} ~.
\]
The idea is that the translation of a
decorated variable $x^r$ is simply the region $r$ so that in the previous
case we obtain the following term of the intuitionistic system:
\[
r_{1} \mid (\lambda x_2.r_2)(\ldots) \mid 
\lambda x_3.r_{3} \mid r_{4} ~.
\]
Note that in the translation the $\nu$'s disappear while the 
$\lambda$'s and $\s{let}$'s are simulated by the intuitionistic
$\lambda$'s. 

Assuming the decoration phase, the forgetful translation $(\forget{~})$ is defined
in Table \ref{forget-translation}.

\begin{lemma} 
The forgetful translation preserves provability
in the following sense:

\begin{enumerate}

\item If $R\AIGives$ then $\forget{R}\IGives$.

\item If $R\AIGives \alpha$ then $\forget{R} \IGives \forget{\alpha}$.

\item If $R\AIGives (\alpha,e)$ then $\forget{R} \IGives (\forget{\alpha},e)$.

\item If $R\AIGives \alpha \leq \alpha'$ then
$\forget{R} \IGives \forget{\alpha}\leq \forget{\alpha'}$.

\item If $R\AIGives (\alpha,e)\leq (\alpha',e')$ then
$\forget{R} \IGives (\forget{\alpha},e) \leq (\forget{\alpha'},e')$.

\item If $R\AIGives \Gamma$ 
then $\forget{R} \AIGives \forget{\Gamma}$.

\item If $R;\Gamma \AIGives P:(\alpha,e)$ (and $P$ has
been decorated) then
$\forget{R};\forget{\Gamma} \IGives \forget{P}:(\forget{\alpha},e)$.
\end{enumerate}
\end{lemma}
\begin{proof}
By induction on the provability relation $\AIGives$.

Concerning the rules for types and region contexts formation and for subtyping,
the forgetful translation provides a one-to-one mapping 
from the rules of the affine-intuitionistic system to the
rules of the intuitionistic one (the only exception are the rules
for $!A$ which become trivial in the intuitionistic framework).
Also note that $\w{dom}(R)= \w{dom}(\forget{R})$. 
With these remarks in mind, the proof of (1-5) is 
straightforward.

The proof of (6) follows directly from (2). We just
notice that the forgetful translation of a context $\Gamma$ 
eliminates all the variable associated with  region types.
The point is that if these variables occur in the program
they will decorated and therefore in the translation they will be replaced
by regions, {\em i.e.}, constants.

In the proof of (7), it is useful to make a few preliminary remarks.
First, {\em weakening} is a derived rule for the intuitionistic system,
so that if we can prove $R;\Gamma\IGives P:(\alpha,e)$ and 
$R,R'\Gives \Gamma,\Gamma'$ then we can
prove $R,R';\Gamma,\Gamma'\IGives P:(\alpha,e)$ too.
Second, if $R_1\csum R_2$ is defined then 
$\forget{R_{1}} = \forget{R_{2}}=\forget{R_{1}\csum R_{2}}$.
The proof is then a rather direct induction on the provability
relation $\AIGives$. 
When we discharge an assumption and when we introduce a formal
parameter with $\lambda$ or with $\s{let}$ we must distinguish
the situation where the variable under consideration has region
type, say, $\regtype{r}{A}$. In this case the variable does not occur in the
translation of the related context $\forget{\Gamma}$ and
it is replaced in the term by the region $r$.
\end{proof}

Next we want to relate the reduction of a program and of its
translation. As already mentioned, in the intuitionistic system all stores are persistent.
Consequently, a reduction such as:
\[
\get{x^r} \mid \store{x^r}{V} \arrow V
\]
might be simulated by 
\[
\get{r} \mid \pstore{r}{\forget{V}} \arrow \forget{V} \mid \pstore{r}{\forget{V}}~.
\]
In other terms, the translated program may contain more values in
the store than the source program. 
To account for this, we introduce a `simulation' relation $\cl{S}$ indexed on  a pair
$R;\Gamma$ such that $R\Gives \Gamma$ and $\Gamma$ is just composed of
variables of region type:
\begin{align*}
\cl{S}_{R;\Gamma}=
\qquad \set{(P,Q) \mid &R;\Gamma \AIGives P:(\alpha,e),\\
                &\forget{R};\_ \IGives Q:(\forget{\alpha},e),\\
                &Q\equiv (\forget{P}\mid S)}  
\end{align*}

\begin{lemma}[simulation]\label{simulation-lemma}
If $(P,Q)\in \cl{S}_{R;\Gamma}$ and $P\arrow P'$ then
$Q\arrow Q'$ and $(P',Q')\in \cl{S}_{R;\Gamma}$.
\end{lemma}
\begin{proof} 
Suppose $(P,Q)\in \cl{S}_{R;\Gamma}$.
Then $(P,\forget{P}) \in \cl{S}_{R;\Gamma}$.
Also if $P\arrow P'$ then $R;\Gamma \AIGives P'$ by 
subject reduction of the affine-intuitionistic system
(incidentally, subject reduction holds for the intuitionistic 
system too \cite{Amadio09}).

By definition $P\arrow P'$ means that $P$ is structurally
equivalent to a process $P_1$ which can be decomposed 
in a static context $C$ and a {\em redex} $\Delta$ of the
shape described in Table \ref{op-sem}.

We notice that the forgetful translation preserves structural equivalence,
namely if $P\equiv P_1$ then $\forget{P}\equiv \forget{P_{1}}$.
Indeed, the commutativity and associativity rules of the affine-intuitionistic
system match those of the intuitionistic system while the rules for commuting
the $\nu$'s are `absorbed' by the translation. For instance,
$\forget{\new{x}{P}\mid P'} = \forget{P}\mid \forget{P'} = \forget{\new{x}{(P\mid P')}}$
with $x$ not free in $P'$.

We also remark that the forgetful translation can be extended to static and evaluation contexts
simply by defining $\forget{[~]}=[~]$. Then we note that
the translation of a static (evaluation) context is an 
intuitionistic static (evaluation) context. In particular, this holds
because the translation of a value is still a value.

Following these remarks, we can derive that $Q\equiv \forget{C}[\forget{\Delta}] \mid S$.
Thus it is enough to focus on the redexes $\Delta$ and show that each reduction
in the affine-intuitionistic system is mapped to a reduction in the intuitionistic one 
and that the resulting program is still related to the program $P'$ via the
relation $\cl{S}_{R;\Gamma}$. 

To this end, we notice that the translation commutes with the substitution
so that $\forget{[V/x]M} = [\forget{V}/x]\forget{M}$. This is a standard argument,
modulo the fact that the variable of region type have to be given a special treatment.
For instance, we have:
\[
\forget{[y^r/x^r]x^r} = \forget{y^r} = r = [r/x^r]r = [\forget{y^r}/x^r]\forget{x^r}~.
\]
Then one proceeds by case analysis on the redex $\Delta$.
Let us look at two cases in some detail.
If $$\Delta = E[\letm{x}{V}{M}] \arrow E[[V/x]M]$$ then 
\[
\begin{array}{rcl}
\forget{\Delta}  &=& \forget{E}[\forget{\letm{x}{V}{M}}]\\
&=& \forget{E}[(\lambda x.\forget{M})\forget{V}]\\
& \arrow& \forget{E}[[\forget{V}/x]\forget{M}]\\
                 &=& \forget{E}[\forget{[V/x]M}] \\
                 &=& \forget{E[[V/x]M}~.
\end{array}
\]
On the other hand if $\Delta = E[\get{x^r}] \mid \store{x^r}{V}$ then
\[
\begin{array}{rcl}
\forget{\Delta}  &=& \forget{E}[\get{r}] \mid \pstore{r}{\forget{V}} \\
                 &\arrow& \forget{E}[V] \mid \pstore{r}{\forget{V}}\\
                 &=&\forget{E[V]} \mid \pstore{r}{\forget{V}}~.
\end{array}
\]
Notice that in this case we have an additional store $\pstore{r}{\forget{V}}$ which
is the reason why in the definition of the relation $\cl{S}$ 
we relate a program to its translation in parallel with some additional store.
\end{proof}

\begin{theorem}[\cite{Amadio09}]\label{thm-ter-intuitionistic}
If $R;\_\IGives P:(\alpha,e)$ then all reductions starting from $P$ terminate.
\end{theorem}

\begin{corollary}[termination]
If $R;\Gamma \AIGives P:(\alpha,e)$ then all reductions starting from $P$ terminate.
\end{corollary}
\begin{proof}
By contradiction. 
We have $(P,\forget{P})\in \cl{S}_{R;\Gamma}$ and 
$R;\_ \IGives \forget{P}:(\forget{\alpha},e)$.
If there is an infinite reduction starting from $P$
then the simulation lemma \ref{simulation-lemma}
entails that there is an infinite reduction starting
form $\forget{P}$. 
And this contradicts the termination of the intuitionistic system
(Theorem \ref{thm-ter-intuitionistic}). 
\end{proof}

 \end{document}